\documentclass[11pt,a4paper]{article}
\usepackage{jheppub}
\usepackage[dvipsnames]{xcolor}
\usepackage{hyperref}
\hypersetup{
 colorlinks,
 citecolor=blue,
 linkcolor=magenta,
 urlcolor=blue}
\usepackage[T1]{fontenc}
\usepackage[utf8]{inputenc}
\usepackage{braket}
\usepackage{float}
\usepackage{cancel}
\DeclareMathAlphabet{\mathbbold}{U}{bbold}{m}{n}
\usepackage{parskip}
\usepackage{xfrac}
\numberwithin{equation}{section}
\usepackage{setspace}
\usepackage{stmaryrd}
\usepackage{youngtab, ytableau}
\Yboxdim{8pt}
\ytableausetup{smalltableaux, boxsize=0.4em}
\def\Xint#1{\mathchoice
{\XXint\displaystyle\textstyle{#1}}%
{\XXint\textstyle\scriptstyle{#1}}%
{\XXint\scriptstyle\scriptscriptstyle{#1}}%
{\XXint\scriptscriptstyle\scriptscriptstyle{#1}}%
\!\int}
\def\XXint#1#2#3{{\setbox0=\hbox{$#1{#2#3}{\int}$ }
\vcenter{\hbox{$#2#3$ }}\kern-.6\wd0}}

\def\dashint{\Xint-}
\usepackage{natbib}
\title{{\bfseries{Large-$N$ Torus Knots in Lens Spaces and Their Quiver Structure}}}
\author{Ritabrata Bhattacharya,}
\author{Suvankar Dutta,}
\author{Naman Pasari,}
\author{Nitin Verma}
\affiliation[a]{\small Department of Physics, Indian Institute of Science Education and Research Bhopal, Bhopal Bypass Road, Bhopal - 462066, India}

\emailAdd{ritabratab3@gmail.com, suvankar@iiserb.ac.in, pasari21@iiserb.ac.in, inittinv@gmail.com}
\emailAdd{}

\abstract{We study torus knot invariants in the lens space $S^{3}/\mathbb{Z}_{p}$ within Chern--Simons theory. Using the surgery and modular description of lens spaces, we derive a general expression for the invariant of an $(\alpha,\beta)$ torus knot in this background. In the large-$N$ limit these invariants simplify and acquire a universal form: the invariant of an $(\alpha,\beta)$ torus knot in $S^{3}/\mathbb{Z}_{p}$ can be expressed in terms of the invariant of the $(\alpha,\alpha+p\beta)$ torus knot in $S^{3}$. After an appropriate redefinition of knot variables, the generating functions of these invariants exhibit a structure analogous to quiver partition functions. Since the associated quiver is independent of the rank $N$ and level $k$ of Chern--Simons theory, the large-$N$ result provides a direct way to identify the underlying quiver, allowing us to determine the quiver structure associated with torus knots in $S^{3}/\mathbb{Z}_{p}$.}

\begin{document}
\maketitle
\flushbottom

\section{Introduction}\label{sec:intro}

Knot invariants in three--manifolds provide a rich framework for exploring the interplay between topology, quantum field theory, and algebraic structures underlying quantum knot invariants. The subject has a long history, beginning with classical polynomial invariants such as the Alexander and Jones polynomials \cite{alexander1928topological,jones1985polynomial,kauffman1990invariant}. A major conceptual advance came with Witten's realization that these invariants arise naturally as Wilson loop observables in three--dimensional Chern--Simons gauge theory \cite{Witten:1988hf}, leading to a deep connection between knot theory, quantum groups, and representation theory \cite{Rosso:1993vn,Labastida:1990bt,Kaul:1991np,RamaDevi:1992np,Nawata:2013qpa}. While the case of $S^{3}$ has been extensively studied and is by now well understood, considerably less is known about knot invariants in more general three--manifolds. Among such backgrounds, lens spaces $S^{3}/\mathbb{Z}_{p}$ provide a natural and tractable extension of the three--sphere, whose nontrivial topology leads to new structures while still allowing explicit computations through the surgery and modular formulation of Chern--Simons theory \cite{Reshetikhin:1991tc,Rozansky_1996,Stevan:2013tha,Rosso:1993vn,Labastida:1990bt,Stevan:2010jh}. Understanding knot invariants in these spaces is therefore an important step toward extending large--$N$ dualities, integrality structures, and the knot--quiver correspondence beyond the simplest $S^{3}$ setting \cite{marino2001framed, labastida2001knots, Liu:2007kv, Kucharski:2017ogk, Kucharski:2025tqb}.

In this work we study torus knots in the lens space $S^{3}/\mathbb{Z}_{p}$ from the perspective of Chern--Simons theory. Using the surgery description of lens spaces and their modular properties, we derive a generic expression for the invariant of an $(\alpha,\beta)$ torus knot in this background. We show that in the double scaling limit
\begin{equation}\label{eq:DSlimit}
    N\to\infty,\; k\to\infty,\quad \frac{N}{k+N} = \lambda\ \text{(fixed)}.
\end{equation}
these invariants admit a particularly simple universal form: the invariant of an $(\alpha,\beta)$ torus knot in $S^{3}/\mathbb{Z}_{p}$ can be expressed directly in terms of the invariant of an $(\alpha,\alpha+p\beta)$ torus knot in $S^{3}$. This relation provides a transparent mapping between different torus knot sectors and isolates the effect of the $\mathbb{Z}_{p}$ quotient in the large--$N$ regime\footnote{Torus knots in lens spaces were studied in \cite{Stevan:2013tha}, where the corresponding knot invariant was decomposed into several components labeled by partitions. It was shown there that the lowest component of any torus knot invariant in the lens space can be expressed as the invariant of a torus knot with shifted parameters in $S^3$. In contrast, we show that such a relation holds for the full knot invariant in the double scaling limit.}. The large--$N$ analysis is performed using matrix model techniques that have proven effective in the study of Chern--Simons knot invariants and torus links \cite{Brini:2011wi,Maji:2023zio}.

An important consequence of this result is that, in the large $N$ limit, the generating functions of torus knot invariants in $S^{3}/\mathbb{Z}_{p}$ acquire a structure closely analogous to quiver partition functions after an appropriate redefinition of the knot variables. This structure allows us to identify the corresponding quiver description for torus knots in lens spaces, thereby extending the knot--quiver correspondence \cite{Kucharski:2017ogk, Ekholm2020, Kucharski:2025tqb, Stokman:1990pa} to the background $S^{3}/\mathbb{Z}_{p}$. A crucial ingredient in this construction is the independence of the quiver structure from the rank and level of the underlying Chern--Simons gauge theory \cite{Kucharski:2017ogk, Ekholm2020}.

To implement this program, we employ matrix model techniques to analyze HOMFLY--PT invariants of torus knots in $S^{3}/\mathbb{Z}_{p}$ in the double-scaling limit (\ref{eq:DSlimit}). Such matrix model descriptions of Chern--Simons theory on lens spaces and related Seifert manifolds provide an efficient framework for large-$N$ computations of correlators and knot observables \cite{Aganagic:2002mv,Halmagyi:2003qt,Chakraborty:2021oyq,Chakraborty:2021bxe,marino2001framed}. In particular, matrix model methods have been successfully applied to torus-knot/torus-link HOMFLY-type invariants in the large-$N$ regime \cite{Brini:2011wi,Maji:2023zio,Stevan:2013tha}.
The resulting large-$N$ relation then allows us to identify the quiver associated with an $(\alpha,\beta)$ torus knot in $S^{3}/\mathbb{Z}_{p}$ in terms of the quiver corresponding to the $(\alpha,\alpha+p\beta)$ torus knot in $S^{3}$. 
Since the quiver structure is independent of the rank $N$ and level $k$ of Chern--Simons theory, this construction determines the quiver partition function in the same double-scaling limit after an appropriate redefinition of the knot variables \cite{Kucharski:2017ogk, Kucharski:2025tqb, Ekholm2020}. 
Consequently, HOMFLY--PT invariants for torus knots in $S^{3}/\mathbb{Z}_{p}$ can be obtained directly from the corresponding quiver data in this limit.

The paper is organized as follows. Section~\ref{sec:knotS3} reviews torus knot invariants in $S^3$ in Chern--Simons theory and their formulation in terms of modular transformations on the torus Hilbert space. Section~\ref{sec:knotlensspace} extends this construction to lens spaces and derives a general expression for torus knot invariants in $S^3/\mathbb{Z}_p$. Section~\ref{sec:matrixmodel} studies their double-scaling limit using matrix model techniques, and section~\ref{sec:reducedinvariant} introduces reduced invariants. Section~\ref{sec:quiver} analyzes the generating functions of these invariants and determines the associated quiver structure. Finally, section~\ref{sec:conclusion} summarizes our results and discusses future directions.

\section{Knot invariants in \(S^3\)}\label{sec:knotS3}

Canonical quantisation of CS theory on a three--manifold $\mathcal{X}$ with boundary $\Sigma$ associates a physical Hilbert space $\mathcal{H}(\Sigma)$ to the boundary. As shown by Witten \cite{Witten:1988hf}, this Hilbert space can be identified with the space of conformal blocks of a Wess--Zumino--Witten (WZW) model on $\Sigma$, with gauge group $G$ at level $k$. The structure of $\mathcal{H}(\Sigma)$ depends strongly on the topology of $\Sigma$. For instance, when $\Sigma=S^2$ the space of conformal blocks is one--dimensional, whereas for $\Sigma=\mathbb{T}^2$ the Hilbert space is nontrivial and admits a canonical basis.

For $\Sigma=\mathbb{T}^2$, the conformal blocks are in one--to--one correspondence with the integrable representations of the affine algebra $\mathbf{g}_k$. A convenient basis of $\mathcal{H}(\mathbb{T}^2)$ is given by vectors $\ket{\mu}$, each labeled by an integrable representation $\mu$. These states arise as the path integral of CS theory on a solid torus with a Wilson loop in representation $\mu$ wrapping the non--contractible cycle. For $G=SU(N)$ or \(U(N)\), the allowed representations correspond to Young diagrams with less than $N$ rows and at most $k$ columns (for \(SU(N)\)) or Young diagrams with negative number of boxes and with a maximum width \(k\) for (\(U(N)\)), ensuring that $\mathcal{H}(\mathbb{T}^2)$ is finite dimensional \cite{Naculich:2007nc}.

Consider a closed three--manifold $\mathcal{M}$ decomposed into two pieces $\mathcal{X}_L$ and $\mathcal{X}_R$ sharing a common boundary $\Sigma$. The path integral on $\mathcal{X}_R$ defines a state $\ket{\phi}\in\mathcal{H}(\Sigma)$, while the path integral on $\mathcal{X}_L$, whose boundary has the opposite orientation, gives a dual state $\bra{\psi}\in\mathcal{H}^*(\Sigma)$. Gluing the two pieces along $\Sigma$ yields the partition function
\begin{equation}
\mathcal{Z}(\mathcal{M})
= \langle \psi | \phi \rangle .
\end{equation}

\begin{figure}[h!]
	\centering
	\includegraphics[width=0.5\textwidth]{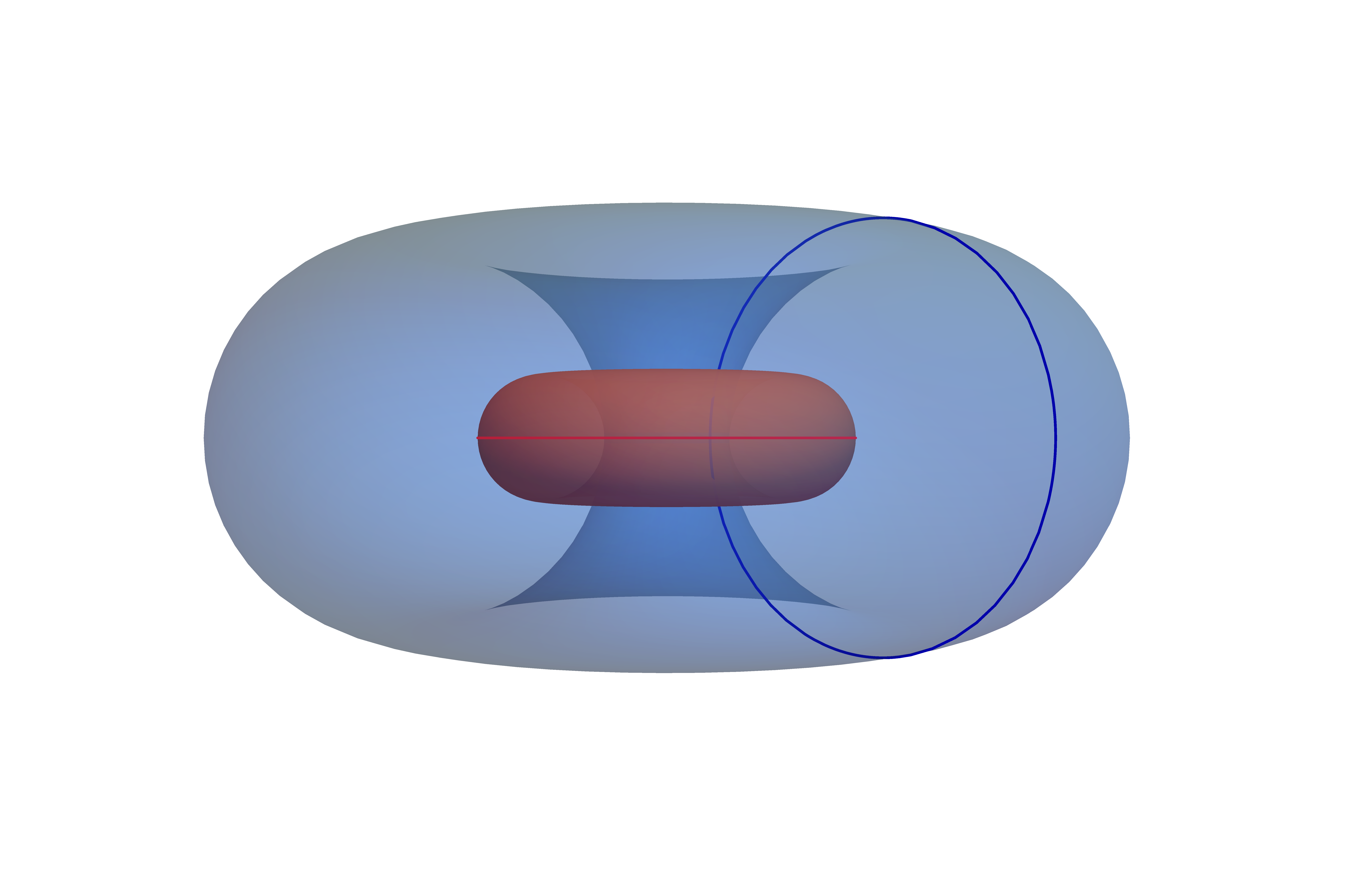}
	\caption{Heegaard Splitting of $S^{3}$. (Credit : \emph{Mathematica})}
	\label{fig:Heegaard Splitting of S3}
\end{figure}

Let us consider $\mathcal{M}=S^3$. Removing a solid torus from $S^3$ leaves a complementary manifold that is again a solid torus. These two solid tori are related by an inversion: the interior of the excised torus becomes the exterior of the complementary one, and vice versa (see fig.~\ref{fig:Heegaard Splitting of S3}). Equivalently, the contractible cycle $\llbracket m \rrbracket$ of one torus is mapped to the non--contractible cycle $\llbracket l \rrbracket$ of the other. Thus $S^3$ can be reconstructed by gluing two solid tori along their boundaries after inverting one of them.
This inversion exchanges the meridian $\llbracket m \rrbracket$ and longitude $\llbracket l \rrbracket$ of the boundary $\mathbb{T}^2$ and is implemented on the Hilbert space $\mathcal{H}(\mathbb{T}^2)$ by the operator $\mathcal{S}$. For further details, see appendix~\ref{app:lens-spac}. Gluing the two solid tori after this operation produces the three--sphere $S^3$. In contrast, gluing two solid tori without this inversion yields the manifold $S^2 \times S^1$.

Applying this construction to Chern--Simons theory, consider the path integral on a solid torus without any Wilson line insertion. The corresponding state in the Hilbert space $\mathcal{H}(\mathbb{T}^2)$ is the vacuum state $\ket{0}$. If we now glue two such solid tori after performing an $\mathcal{S}$ modular transformation on the boundary of one of them, the resulting partition function is given by
\begin{equation}\label{eq:S3pf}
\mathcal{Z}(S^3) = \bra{0}\mathcal{S}\ket{0} = \mathcal{S}_{00}.
\end{equation}
This provides the standard expression for the Chern--Simons partition function on the three--sphere. 

Next, consider a Wilson loop in representation $\mu$ supported on the non--contractible cycle (the longitude) of a solid torus. The path integral on the torus then prepares the state $\ket{\mu} \in \mathcal{H}(\mathbb{T}^2)$. We now glue this torus to another solid torus without any Wilson line insertion. As discussed earlier, the gluing required to produce $S^3$ involves a modular $\mathcal{S}$ transformation acting on the boundary of one torus. The resulting partition function computes the Chern--Simons invariant of an unknot in representation $\mu$:
\begin{equation}
W(\text{unknot}; \mu)
= \bra{0} \mathcal{S} \ket{\mu}
= \mathcal{S}_{0\mu}.
\end{equation}
Geometrically, the unknot may be characterised by its winding numbers around the cycles of the boundary torus. With respect to the torus on which it is initially defined, the unknot wraps once around the longitude and does not wind along the meridian. It may therefore be represented as
\begin{equation}
(1,0) \equiv 1\llbracket l \rrbracket + 0 \llbracket m \rrbracket 
\end{equation}

However, the modular $\mathcal{S}$ transformation exchanges the meridian and longitude cycles. Consequently, after gluing, the same knot can equivalently be viewed as
\begin{equation}
(0,1) \equiv 0\llbracket l \rrbracket + 1\llbracket m \rrbracket 
\end{equation}
with respect to the complementary solid torus. Since knot invariants are independent of this description, we obtain
\begin{equation}
W^{(1,0)}_\mu(S^3) = W^{(0,1)}_\mu(S^3) = \mathcal{S}_{0\mu}.
\end{equation}

We now consider a general torus knot embedded in a solid torus. An $(\alpha,\beta)$ - torus knot, $\mathbb T^{(\alpha,\beta)}$ is specified up to isotopy by the homology class 
\begin{equation}
(\alpha,\beta) =  \alpha \llbracket l \rrbracket + \beta \llbracket m \rrbracket.
\end{equation}
such that $\gcd(\alpha, \beta) = 1$. By the above Definition, a torus knot $\mathbb{T}^{(\alpha,\beta)}$ can be intuitively thought of as a closed loop that wraps $\alpha$ times along the longitudinal direction and $\beta$ times around the meridonial direction. An example has been illustrated in the fig. \ref{fig:Trefoil Wrapped on Torus} below,
\begin{figure}[h!]
	\centering
	\includegraphics[width=0.5\textwidth]{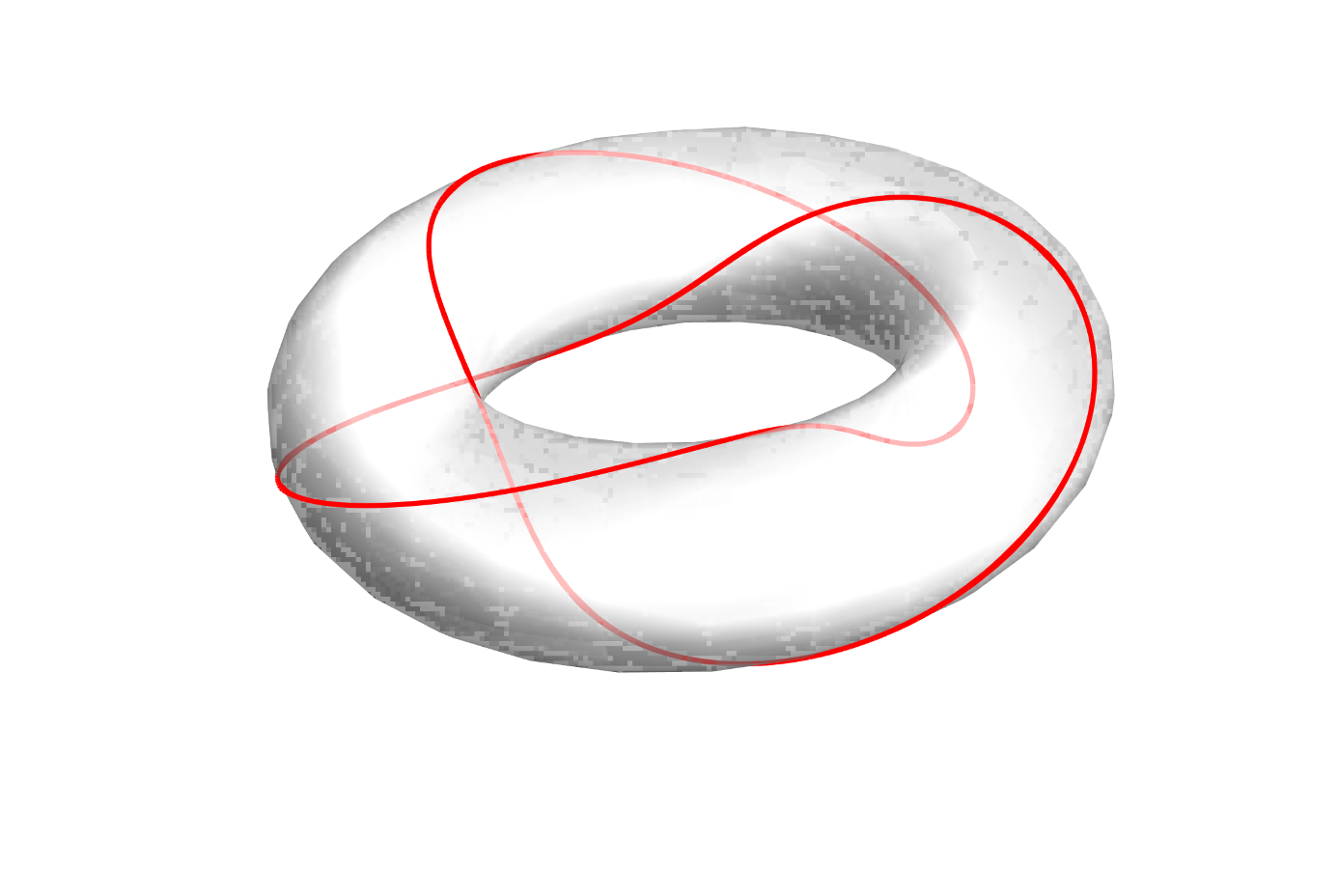}
	\caption{Trefoil $\mathbb T^{(2,3)}$, wrapped around the torus. (Credit-\emph{Mathematica})}
	\label{fig:Trefoil Wrapped on Torus}
\end{figure}

The path integral on the solid torus in the presence of a Wilson loop in
representation $\mu$ wrapping the $(\alpha,\beta)$ cycle prepares a state
in the Hilbert space $\mathcal{H}(\mathbb{T}^2)$.
This state, often referred to as a \emph{knot state}, may be expanded in the
standard representation basis as
\cite{Labastida:1990bt,Kaul:1991np,Stevan:2010jh,Stevan:2013tha,Stevan:2014xma,labastida2001knots,Labastida:1995ki}.
\begin{eqnarray}
\ket{\mathbb{T}^{(\alpha,\beta)};\mu,f}
=
q^{(f- \alpha \beta)\kappa_{\mu}}\sum_{\rho \atop |\rho| = \alpha |\mu|}
c^{\rho}_{\mu,\alpha}
\, q^{\frac{\beta}{\alpha}\kappa_{\rho} }
\ket{\rho}, \quad \alpha\neq 0.
\label{eq:Stevan20}
\end{eqnarray}
Here we define the knot states with an overall phase factor \(q^{(f-\alpha\beta)\kappa_\mu}\). Later we see that this extra factor plays the role of framing. $c^{\rho}_{\mu,\alpha}$ denotes the Adams coefficients. In Chern--Simons theory these coefficients determine the decomposition of Wilson loop operators,
\begin{equation}\label{eq:adam-coefficients}
\mathrm{Tr}_\mu(U^\alpha)
=
\sum_{\rho}
c^{\rho}_{\mu,\alpha} \,
\mathrm{Tr}_\rho(U).
\end{equation}
The parameter \(q\) is given by
\begin{eqnarray}
    q = \exp\left(\frac{2\pi i}{k + N}\right).
\end{eqnarray}
We also use another variable \(v\) to express the HOMFLY-PT polynomials given by,
\begin{equation}
    v = q^N = \exp\left(\frac{2\pi i N}{k + N}\right).
\end{equation}
The quantity $\kappa_\rho$ is defined in terms of the quadratic Casimir associated with the representation $\rho$
\begin{equation}\label{eq:kappadef}
    \kappa_{\rho} = \frac{1}{2}\sum_{i=1}^Nl_i(l_i-2i+1)+\frac{N|\rho|}{2}.
\end{equation}
\(l_i\) is the number of boxes in the i-th row in the Young diagram associated with the representation \(\rho\) and \(|\rho|\) is the number of boxes in the Young diagram. The notation
\[
\sum_{\rho \atop |\rho| = \alpha |\mu|}
\]
denotes the sum over all representations $\rho$ whose total number of boxes equals $\alpha |\mu|$.

The states $\ket{\rho}$ appearing on the right--hand side of the knot state (\ref{eq:Stevan20}) correspond to Young diagram states associated with partitions of $\alpha |\mu|$, and are not, strictly speaking, elements of the Hilbert space $\mathcal{H}(\mathbb{T}^2)$. However, in the combined large $N$ and large $k$ limit, this distinction becomes immaterial, and one may consistently regard the states $\ket{\rho}$ as belonging to $\mathcal{H}(\mathbb{T}^2)$. With this understanding, the knot state construction shows that a general $(\alpha,\beta)$ torus knot state admits a representation as a linear superposition of unknot states labelled by different representations. In other words, arbitrary torus knot states can be decomposed into sums over Wilson loop states associated with unknots in various representations.

To compute the torus knot invariant in $S^3$, we glue this solid torus to an empty solid torus. As discussed earlier, the gluing that produces $S^3$ requires the insertion of the modular $\mathcal{S}$ transformation. The resulting amplitude is
\begin{equation}
\bra{0}\mathcal{S}\ket{\mathbb{T}^{(\alpha,\beta)};\mu,f}
=
q^{(f- \alpha \beta)\kappa_{\mu}}\sum_{\substack{\rho \\ |\rho| = \alpha |\mu|}}
c^{\rho}_{\mu,\alpha}
\, q^{\frac{\beta}{\alpha}\kappa_{\rho} }
\bra{0}\mathcal{S}\ket{\rho}.
\end{equation}
Using the definition of the modular matrix elements,
\begin{equation}
\bra{0}\mathcal{S}\ket{\rho} = \mathcal{S}_{0\rho},
\end{equation}
we obtain
\begin{equation}
\bra{0}\mathcal{S}\ket{\mathbb{T}^{(\alpha,\beta)};\mu,f}
=
q^{(f- \alpha \beta)\kappa_{\mu}}\sum_{\substack{\rho \\ |\rho| = \alpha |\mu|}}
c^{\rho}_{\mu,\alpha}
\, q^{\frac{\beta}{\alpha}\kappa_{\rho} }
\mathcal{S}_{0\rho}.
\end{equation}

The normalized torus knot invariant is obtained by dividing by the vacuum partition function on $S^3$,
\begin{equation}
Z(S^3) = \bra{0}\mathcal{S}\ket{0} = \mathcal{S}_{00}.
\end{equation}
Thus,
\begin{equation}
W^{(\alpha,\beta)}_\mu(S^3,f;q,v)
=
\frac{\bra{0}\mathcal{S}\ket{\mathbb{T}^{(\alpha,\beta)};\mu,f}}
{\bra{0}\mathcal{S}\ket{0}}
=
q^{(f- \alpha \beta)\kappa_{\mu}}\sum_{\substack{\rho \\ |\rho| = \alpha |\mu|}}
c^{\rho}_{\mu,\alpha}
\, q^{\frac{\beta}{\alpha}\kappa_{\rho} }
\frac{\mathcal{S}_{0\rho}}{\mathcal{S}_{00}}.
\end{equation}
Using the standard identity
\begin{equation}\label{eq:dimqdef}
\frac{\mathcal{S}_{0\rho}}{\mathcal{S}_{00}} = \dim_q(\rho),
\end{equation}
we arrive at the final expression
\begin{equation}\label{eq:torusknotS3}
W^{(\alpha,\beta)}_\mu(S^3,f;q,v)
=
q^{(f- \alpha \beta)\kappa_{\mu}}\sum_{\substack{\rho \\ |\rho| = \alpha |\mu|}}
c^{\rho}_{\mu,\alpha}
\, q^{\frac{\beta}{\alpha}\kappa_{\rho} }
\, \dim_q(\rho).
\end{equation}
This expression was earlier derived in \cite{RamaDevi:1992np, Ramadevi:1994zb,  Maji:2023zio}.

For unknot \((1,0)\), the Adams coefficients $c^\rho_{\mu,1}$ are given by
\begin{equation}
c^\rho_{\mu,1} = \delta^\rho_\mu ,
\end{equation}
which follows directly from their defining relation (\ref{eq:adam-coefficients}). As a consistency check, for the unknot $(1,0)$ one recovers the familiar result (with framing \(f=0\))
\begin{equation}
W_\mu^{(\text{unknot})}(S^3,f=0;q,v)
=
\frac{\mathcal{S}_{0\mu}}{\mathcal{S}_{00}}
=
\dim_q(\mu),
\end{equation}
which expresses the unknot invariant in terms of the quantum dimension of the representation $\mu$.

Consequently, the $(\alpha,\beta)$ torus knot invariant in $S^3$ may be expressed as a sum over $(1,0)$ unknot invariants in $S^3$. In other words, the general torus knot invariant admits a decomposition in terms of contributions associated with unknots in various representations.

As discussed earlier, the modular $\mathcal{S}$ transformation exchanges the meridian and longitude cycles of the boundary torus. Consequently, a torus knot characterised by winding numbers $(\alpha,\beta)$ with respect to one solid torus is equivalently described as a $(\beta,\alpha)$ torus knot with respect to the complementary solid torus.

Since the resulting three--manifold is unchanged and knot invariants are independent of this choice of description, we obtain the symmetry
\begin{equation}
W^{(\alpha,\beta)}_\mu(S^3,f;q,v)
=
W^{(\beta,\alpha)}_\mu(S^3,f;q,v).
\end{equation}

\subsection{\((2,2n+1)\) torus knot invariants in symmetric representation}

We use (\ref{eq:torusknotS3}) to write down the invariants for any \((2,2n+1)\) torus knot invariants in symmetric \(r\) representation, as a special case, that will be used for later purpose. The Adam's coefficients for symmetric representations \(\mu =r\) and \(\alpha = 2\) are non-zero only when the partitions \(\rho\) has maximum two rows with total number of boxes \(2r\). Such partitions can be represented by \((2r-k, k)\) with \(0\leq k \leq r\). The value of Adam's coefficients for such partitions are given by
\begin{equation}
    c^{(2r-k,k)}_{r,2} = (-1)^k .
\end{equation}
Therefore, the \((2, 2n+1)\) torus knot invariants are given by,
\begin{eqnarray}\label{eq:2-2n+1torusknot}
    W^{(2,2n+1)}_r(S^3,f;q,v) = q^{(f-2(2n+1))\kappa_r} \sum_{k = 0}^r (-1)^k\ q^{\frac{2n+1}{2}\kappa_{(2r-k,k)}} \dim_q(2r-k,k).
\end{eqnarray}

\subsection{Torus knot invariants in sheared \(S^3\)}\label{sec:knotshearedS3}

Instead of performing a pure $\mathcal{S}$--modular surgery, one may consider a
more general gluing obtained by combining inversion with a Dehn twist. The Dehn twist along one fundamental cycle of the torus is implemented on $\mathcal{H}(\mathbb{T}^2)$ by the modular $T$-matrix $\mathcal{T}$. A particularly important example is provided by the $\mathcal{TST}$ modular
transformation. When two empty solid tori are glued using $\mathcal{TST}$,
the resulting three--manifold is again $S^3$, although the identification of
cycles differs from that of the simple $\mathcal{S}$--transformation.
Recall that the modular transformation $\mathcal{S}$ exchanges the longitude
and meridian cycles, while $\mathcal{T}$ implements a Dehn twist (appendix \ref{app:lens-spac}).
Under the action of $\mathcal{TST}$, the cycles transform as
\begin{equation}
\llbracket l \rrbracket \;\longrightarrow\; \llbracket l \rrbracket,
\qquad
\llbracket m \rrbracket \;\longrightarrow\; \llbracket m \rrbracket + \llbracket l \rrbracket .
\end{equation}
Thus, the longitude remains unchanged, while the meridian is shifted by the longitude. In contrast to pure $\mathcal{S}$--surgery, the cycles are therefore not simply interchanged. In this description the resulting three--manifold remains $S^3$, however, the $S^3$ is realised as the union of an ordinary solid torus and a complementary solid torus that is not only inverted but also twisted, and may therefore be viewed as a sheared presentation of $S^3$. Applying this construction to Chern--Simons theory, the path integral on an empty solid torus prepares the vacuum state $\ket{0}$. Gluing two such tori using the $\mathcal{TST}$ transformation yields the partition function of Chern--Simons theory in a sheared \(S^3\) 
\cite{Blau:2013oha,Blau:2016vfc}
\begin{equation}\label{eq:TSTpf}
\mathcal{Z}_{\rm{TST}}(S^3)
=
\bra{0}\mathcal{TST}\ket{0}
=
\mathcal{T}_{00} \mathcal{S}_{00}^2.
\end{equation}
Compared to the standard result obtained from pure $\mathcal{S}$--gluing (\ref{eq:S3pf}), the partition function differs by the additional factor $T_{00}^{\,2}$. Since $\mathcal{T}_{00}$ is a pure phase, this difference reflects a change of framing rather than a change of manifold. 

We can also apply the $\mathcal{TST}$ modular surgery to a torus knot state. A subtlety in this construction arises in identifying the torus knot after the gluing. Because the complementary torus is twisted by the modular transformation, its meridian and longitude cycles are mixed, making a direct geometric interpretation of the resulting knot less transparent.

To analyse this situation, we proceed as follows. We consider Chern--Simons theory on two solid tori, labelled as the left and right tori, and insert an $(\alpha,\beta)$ torus knot in the right torus. The path integral on the right torus prepares the corresponding torus knot state, given in (\ref{eq:Stevan20}), while the left torus without Wilson line insertions prepares the vacuum bra $\bra{0}$. We then act with the $\mathcal{TST}$ modular transformation on the left torus and glue it to the right torus.

With this prescription, the torus on which the knot is initially defined remains untwisted, so the knot retains its natural interpretation as an $(\alpha,\beta)$ torus knot with respect to that torus. The complementary torus, however, becomes both inverted and twisted. As a consequence, its meridian and longitude cycles are no longer simply exchanged but are mixed, and the same embedded knot may therefore admit a different description when expressed in terms of the transformed cycles.

For this reason, when we refer to an $(\alpha,\beta)$ torus knot in the sheared $S^3$, we mean that the knot is defined with respect to the untwisted solid torus inside $S^3$. Thus, starting from an $(\alpha,\beta)$ torus knot state, the knot obtained after $\mathcal{TST}$ surgery is naturally interpreted as an $(\alpha,\beta)$ torus knot in a sheared presentation of $S^3$.

Starting from the torus--knot state (\ref{eq:Stevan20}) the (unnormalised) amplitude obtained by $\mathcal{TST}$--gluing is
\begin{equation}
\bra{0}\,\mathcal{TST}\,\ket{\mathbb{T}^{(\alpha,\beta)};\mu,f}
=
q^{(f- \alpha \beta)\kappa_{\mu}}\sum_{\substack{\rho\\ |\rho|=\alpha|\mu|}}
c^{\rho}_{\mu,\alpha}\;
q^{\frac{\beta}{\alpha}\,\kappa_{\rho}}\;
\bra{0}\,\mathcal{TST}\,\ket{\rho}.
\end{equation}
Since $\mathcal{T}$ is diagonal, one has
\begin{equation}
\bra{0}\,\mathcal{TST}\,\ket{\rho}
=
(\mathcal{TST})_{0\rho}
=
\mathcal{T}_{00}\,\mathcal{S}_{0\rho}\,\mathcal{T}_{\rho\rho}.
\end{equation}
The corresponding partition function for this gluing is given by (\ref{eq:TSTpf}). Therefore the normalised knot invariant is given by
\begin{align}
W^{(\alpha,\beta)}_{\mu}(S^3,f;q,v)\big|_{\rm{TST}}
&=
\frac{1}
{\mathcal{Z}_{\rm{TST}}(S^3)} \bra{0}\,\mathcal{TST}\,\ket{\mathbb{T}^{(\alpha,\beta)};\mu}
\nonumber\\[2pt]
&=
q^{(f- \alpha \beta)\kappa_{\mu}} \sum_{\substack{\rho\\ |\rho|=\alpha|\mu|}}
c^{\rho}_{\mu,\alpha}\;
q^{\frac{\beta}{\alpha}\,\kappa_{\rho}}\;
\frac{\mathcal{S}_{0\rho}}{\mathcal{S}_{00}}\;
\frac{\mathcal{T}_{\rho\rho}}{\mathcal{T}_{00}}.
\end{align}
Using (\ref{eq:dimqdef}) and 
\begin{eqnarray}\label{eq:TbyTtoqkappa}
    \frac{\mathcal{T}_{\rho\rho}}{\mathcal{T}_{00}} = q^{\kappa_\rho},
\end{eqnarray}
this can be written as
\begin{equation}\label{eq:WTST}
W^{(\alpha,\beta)}_{\mu}(S^3,f;q,v)\big|_{\rm{TST}}
=
q^{(f- \alpha \beta)\kappa_{\mu}} \sum_{\substack{\rho\\ |\rho|=\alpha|\mu|}}
c^{\rho}_{\mu,\alpha}\;
q^{\frac{\beta+\alpha}{\alpha}\,\kappa_{\rho}}\;
\dim_q(\rho).
\end{equation}

The invariant obtained from $\mathcal{TST}$-surgery should not be interpreted as a mere framing modification. Although the resulting three--manifold is still $S^3$, the gluing now involves an additional Dehn twist, which alters the identification of cycles on the boundary torus. Consequently, the ambient geometry is naturally viewed as a sheared presentation of $S^3$.

In this sheared description, a torus knot labeled by winding numbers $(\alpha,\beta)$ retains its original interpretation with respect to the torus on which it is defined. However, when the same configuration is expressed in the standard (unsheared) presentation of $S^{3}$, the mixing of the longitude and meridian cycles induced by the Dehn twists modifies the slope of the knot. As a consequence, the knot invariant computed using $\mathcal{TST}$-surgery can be naturally related to the invariant obtained in the usual $\mathcal{S}$-surgery description. In particular, one finds
\begin{equation}\label{eq:WTST-WS}
W_\mu^{(\alpha,\beta)}(S^3,f;q,v)\Big|_{\mathrm{TST}}
= q^{\alpha^2 \kappa_\mu}~
W_\mu^{(\alpha,\beta+\alpha)}(S^3,f;q,v).
\end{equation}
Thus, an $(\alpha,\beta)$ torus knot defined in the sheared presentation of $S^{3}$ is equivalent to an $(\alpha,\alpha+\beta)$ torus knot in the standard presentation of $S^{3}$.

\section{Knot invariants in lens space}\label{sec:knotlensspace}

A Lens space $L(p,1) \equiv S^3/\mathbb{Z}_p$ is obtained in the surgery construction by gluing two solid tori with the modular transformation (see appendix \ref{app:lens-spac}) \cite{Blau:2013oha, Blau:2016vfc}
\begin{equation}
\mathcal{K} = (\mathcal{TST})^{p}.
\end{equation}
Under the modular transformation $(\mathcal{TST})^{p}$, the cycles of the boundary torus transform as
\begin{equation}
(\mathcal{TST})^{p} :
\qquad
\llbracket l \rrbracket \;\longrightarrow\; \llbracket l \rrbracket,
\qquad
\llbracket m \rrbracket \;\longrightarrow\; \llbracket m \rrbracket + p \,\llbracket l \rrbracket .
\end{equation}
Using the modular group relation
\begin{equation}
\mathcal{TST} = \mathcal{S}^\dagger\mathcal{T}^{-1}\mathcal{S},
\end{equation}
\(\mathcal K\) may equivalently be written as
\begin{equation}
\mathcal{K} = \mathcal{S}^\dagger\mathcal{T}^{-p}\mathcal{S}.
\end{equation}
Thus, the Lens space $S^3/\mathbb{Z}_p$ is constructed by gluing two solid tori via the transformation $(\mathcal{TST})^{p}$. In the Chern--Simons formulation, the corresponding partition function is given by
\begin{equation}\label{eq:ZLS}
\mathcal Z\big(S^3/\mathbb{Z}_p\big) = 
\bra{0} \mathcal{K}\ket{0}=
\sum_\rho \mathcal{S}_{0\rho}^2 \mathcal{T}_{\rho\rho}^{-p}.
\end{equation}
Geometrically, the repeated $\mathcal{T}$ insertions implement Dehn twists, while $\mathcal{S}$ exchanges the meridian and longitude cycles, thereby generating the nontrivial topology of the Lens space.

\subsection{\((\alpha, \beta)\) torus knots in lens space}\label{sec:torusknotlensspace}

To compute the torus knot invariant in $S^3/\mathbb{Z}_p$, we begin with two solid tori. The left solid torus is taken to be empty and therefore prepares the vacuum state $\bra{0}$. On the right solid torus we insert a $(\alpha,\beta)$ torus knot, and the corresponding state is given by (\ref{eq:Stevan20}). 

In order to generate the torus knot invariants in Lens space, we apply the modular transformation
\begin{equation}
\mathcal{K} = (\mathcal{TST})^{p}
\end{equation}
to one of the solid tori before gluing. For definiteness, let us perform this operation on the left solid torus. Under this transformation, the meridian and longitude cycles are modified. Gluing the transformed left torus with the right torus then produces the manifold $S^3/\mathbb{Z}_p$ in the presence of a torus knot.

In this construction, the right solid torus remains untwisted and naturally defines an embedded torus inside the resulting manifold. Consequently, the Wilson loop retains its interpretation as a $(\alpha,\beta)$ torus knot with respect to this untwisted torus, while the nontrivial topology of the ambient space is entirely encoded in the modular transformation acting on the complementary torus.

Starting from the torus knot state (\ref{eq:Stevan20}) the normalized torus knot invariant in the Lens space is defined by
\begin{equation}
W^{(\alpha,\beta)}_\mu(S^3/\mathbb{Z}_p,f;q,v)
= \frac{1}{\mathcal{Z}(S^3/\mathbb{Z}_p)}\bra{0}\mathcal{S}^\dagger \mathcal{T}^{-p}\mathcal{S}
\ket{\mathbb{T}^{(\alpha,\beta)};\mu}.
\end{equation}
Substituting the knot state, we obtain
\begin{equation}
W^{(\alpha,\beta)}_\mu(S^3/\mathbb{Z}_p,f;q,v)
=
q^{(f- \alpha \beta)\kappa_{\mu}} \sum_{\substack{\rho \\ |\rho|=\alpha|\mu|}}
c^\rho_{\mu,\alpha}\;
q^{\frac{\beta}{\alpha}\kappa_\rho}\;
\frac{
(\mathcal{S}^\dagger \mathcal{T}^{-p}\mathcal{S})_{0\rho}
}{
(\mathcal{S}^\dagger \mathcal{T}^{-p}\mathcal{S})_{00}
}.
\end{equation}
Using the unitarity and symmetry of the modular $\mathcal{S}$ matrix,
\begin{equation}
(\mathcal{S}^\dagger)_{0\rho}=\mathcal{S}_{0\rho},
\end{equation}
the matrix element may be written as
\begin{equation}
(\mathcal{S}^\dagger \mathcal{T}^{-p}\mathcal{S})_{0\rho}
=
\sum_{\nu}
\mathcal{S}_{0\nu}\,
\mathcal{T}_{\nu\nu}^{-p}\,
\mathcal{S}_{\nu\rho}
=
\sum_{\nu}
\mathcal{S}_{0\nu}^2\,
\mathcal{T}_{\nu\nu}^{-p}\,
\frac{\mathcal{S}_{\nu\rho}}{\mathcal{S}_{0\nu}}.
\end{equation}
Therefore, the normalized torus knot invariant takes the form
\begin{equation}\label{eq:Wabfp}
W^{(\alpha,\beta)}_\mu(S^3/\mathbb{Z}_p,f;q,v)
=
q^{(f- \alpha \beta)\kappa_{\mu}} \sum_{\substack{\rho \\ |\rho|=\alpha|\mu|}}
c^\rho_{\mu,\alpha}\;
q^{\frac{\beta}{\alpha}\kappa_\rho}\;
\Xi^{(p)}_\rho.
\end{equation}
where,
\begin{equation}\label{eq:Xipl}
    \Xi^{(p)}_\rho = \frac{1}{\mathcal{Z}(S^3/\mathbb{Z}_p)}
\sum_{\nu}
\mathcal{S}_{0\nu}^2\,
\mathcal{T}_{\nu\nu}^{-p}\,
\frac{\mathcal{S}_{\nu\rho}}{\mathcal{S}_{0\nu}}.
\end{equation}
This expression is exact for arbitrary rank $N$ and level $k$ of Chern--Simons theory, but its explicit evaluation is cumbersome since the sum over $\nu$ runs over all integrable representations of the affine algebra. In the double-scaling limit~(\ref{eq:DSlimit}), it simplifies significantly and can be analyzed using matrix model techniques.

\subsection{Matrix model}\label{sec:matrixmodel}

In the large-$N$ limit, Chern--Simons theory admits a matrix model description in which knot invariants arise as operator expectation values \cite{Chakraborty:2021oyq, Chakraborty:2021bxe, Maji:2023zio}. This framework provides an efficient way to analyze the double-scaling behavior of torus knot invariants in lens spaces and uncover their quiver structure.

We begin by recalling the modular transformation matrices of the ${\mathfrak u}(N)_k$ WZW theory. The modular $\mathcal{S}$--matrix is given by \cite{Naculich:2007nc, Naculich:1990pa, MLAWER1991863}
\begin{equation}\label{eq:Sdef}
\mathcal{S}_{\mu\rho}
=
\frac{(-i)^{N(N-1)/2}}{(k+N)^{N/2}}
\exp\!\left[-
\frac{2\pi i\, l^{(\mu)} l^{(\rho)}}{N(N+k)}
\right]
\det M(\mu,\rho),
\end{equation}
where
\begin{equation}
M_{ij}(\mu,\rho)
=
\exp\!\left[
\frac{2\pi i}{k+N}\,
\phi_i(\mu)\phi_j(\rho)
\right],
\qquad i,j=1,\ldots,N,
\end{equation}
and
\begin{equation}
\phi_i(\mu)
=
l_i^{(\mu)}-\frac{l^{(\mu)}}{N}-i+\frac{N+1}{2}.
\end{equation}
Here $l^{(\mu)}$ denotes the total number of boxes in the Young diagram associated with the representation $\mu$, while $l_i^{(\mu)}$ denotes the number of boxes in the i-th row.

The modular $\mathcal{T}$--matrix is diagonal and takes the form \cite{Naculich:2007nc, Naculich:1990pa, MLAWER1991863}
\begin{equation}\label{eq:Tdef}
\mathcal{T}_{\mu\rho}
=
\exp\!\left[
2\pi i \left(
\mathcal{Q}_\mu - \frac{c}{24}
\right)
\right]
\delta_{\mu\rho},
\end{equation}
where
\begin{equation}
\mathcal{Q}_\mu=\frac{\kappa_\mu}{k+N},
\qquad
c=\frac{N(Nk+1)}{N+k}.
\end{equation}
Following \cite{Chakraborty:2021bxe,Maji:2023zio} we parametrise the representation $\rho$ in terms of angular variables $\theta_i$, defined by 
\begin{equation}\label{eq:thetaidef}
\theta_i
=
\frac{2\pi}{N+k}
\left(
l_i -i + \frac{N+1}{2}
\right).
\end{equation}
Since $\rho$ is an integrable representation, \(l_i\) are bounded: \(-\frac{k}{2}\leq l_i\leq \frac{k}{2}\). In the double scaling limit, the variables $\theta_i$ range continuously over the interval $[-\pi,\pi]$, and may therefore be interpreted as eigenvalues of an $N\times N$ unitary matrix,
\begin{equation}
U=\mathrm{diag}(e^{i\theta_1},\ldots,e^{i\theta_N}).
\end{equation}

With this parametrisation, the modular matrix elements admit a natural group--theoretic interpretation. In particular, consider the ratio
\begin{equation}
\frac{\mathcal{S}_{\rho\mu}}{\mathcal{S}_{0\rho}}.
\end{equation}
Substituting the determinant representation of the $\mathcal{S}$--matrix and expressing the result in terms of hook numbers, one finds that all overall normalisation factors cancel. The dependence on the representation $\rho$ then enters solely through the variables $\theta_i$. After straightforward algebra, this ratio reduces to
\begin{equation}
\frac{\mathcal{S}_{\rho\mu}}{\mathcal{S}_{0\rho}}
=
\mathbf{s}_\mu(U),
\end{equation}
where $\mathbf{s}_\mu(U)$ denotes the Schur polynomial,
\begin{equation}
\mathbf{s}_\mu(U)
=
\frac{
\det \left[e^{i\theta_i h_j^{(\mu)}}\right]
}{
\det \left[e^{i\theta_i (N-j)}\right]
}
\end{equation}
where \(h_j^{(\mu)} = l_i^{(\mu)} + N -i\) are the hook numbers associated with the representation \(\mu\). Thus, the modular matrix ratio is identified with the character of the unitary matrix $U$ in representation $\mu$, while the eigenvalues of $U$ are determined by the number of boxes in representation $\rho$ (\ref{eq:thetaidef}).

The summation over representations may be naturally interpreted as a summation over unitary matrices. In the hook-number parametrisation, the factor $\mathcal{S}_{0\rho}^2$ takes a particularly simple form. When expressed in terms of the variables $\theta_i$, it reduces to the familiar Vandermonde measure associated with unitary matrix integrals. Consequently, in the large $N$ limit, the discrete sum over representations can be replaced by a continuous integration over eigenvalues,
\begin{equation}\label{eq:RtoU}
\sum_{\rho} \mathcal{S}_{0\rho}^2
\;\longrightarrow\;
\int \prod_i d\theta_i \prod_{i<j}
\sin^2\!\left(\frac{\theta_i-\theta_j}{2}\right)
\;\equiv\;
\int [DU],
\end{equation}
where $\int[DU]$ denotes the Haar measure of the unitary group. In writing this expression, we have suppressed overall factors independent of the variables $\theta_i$, as they do not influence the subsequent analysis.

With this replacement, the partition function (\ref{eq:ZLS}) admits a representation as a zero-dimensional unitary matrix model with an effective weight factor $e^{-(N+k)^2 S[\theta]}$,
\begin{equation}\label{eq:pfU}
\mathcal{Z}(S^{3}/\mathbb{Z}_{p})
=
\int \prod_i d\theta_i \prod_{i<j}
\sin^2\!\left(\frac{\theta_i-\theta_j}{2}\right)
\, e^{-(N+k)^2 S[\theta]}.
\end{equation}
The corresponding effective action is given by
\begin{equation}\label{eq:weight}
S[\theta]
=
\frac{p\rho}{N\pi}
\sum_{i=1}^N
\left(
\frac{\theta_i^2}{4}-\frac{\pi^2}{12}
\right)
+
\frac{\pi p \lambda(1-\lambda)}{12}.
\end{equation}
In arriving at this form, we have performed the analytic continuation $p \rightarrow -ip$ in (\ref{eq:ZLS}), which is convenient for expressing the matrix model in a standard form
\begin{equation}\label{eq:pfU2}
\mathcal{Z}(S^{3}/\mathbb{Z}_{p})
=
\int \prod_i d\theta_i \exp\left[-N^2 S_{\text{eff}}(\theta_i)\right]
\end{equation}
where,
\begin{equation}\label{eq:Seff}
S_{\text{eff}} = -\frac{1}{N^2} \sum_{i\neq j}
\ln\left|\sin\!\left(\frac{\theta_i-\theta_j}{2}\right)\right|
+ \frac{p}{\lambda \pi}
\frac{1}{N}\sum_{i=1}^N
\left(
\frac{\theta_i^2}{4}-\frac{\pi^2}{12}
\right)
+
\frac{\pi \lambda(1-\lambda)}{12 p}.
\end{equation}
The saddle point equation obtained from the partition function (\ref{eq:pfU}) is given by
    \begin{equation}\label{eq:sad0}
          \dashint  \rho(\theta')\cot\left(\frac{\theta-\theta'}{2}\right)d\theta' = \frac{p}{2\pi\lambda}\theta.
\end{equation}
The solution to this equation is given by\footnote{In the double-scaling limit of $U(N)$ Chern--Simons theory, the integrability constraint on representations imposes an upper bound $1/(2\pi\lambda)$ on the eigenvalue density of the associated matrix model \cite{Chakraborty:2021oyq, Chakraborty:2021bxe}. When the saddle point distribution saturates this bound, the system undergoes a third-order phase transition analogous to the Douglas--Kazakov transition in two-dimensional Yang--Mills theory. In this work, however, we do not analyze this transition and restrict ourselves to computing the invariants in the phase described by the saddle point solution (\ref{eq:bgsol}).},
\begin{align}
    \label{eq:bgsol}
    \rho(\theta)&=\frac{p}{2\pi^2\lambda}\tanh^{-1}\sqrt{1-\frac{e^{-2\pi \lambda/p}}{\cos^{2}(\theta/2)}},\quad \text{with}~-2\sec^{-1}e^{\pi \lambda/p}<\theta < 2 \sec^{-1}e^{\pi \lambda/p}
\end{align}

Converting the sum over integrable representation to integration over unitary matrices,  we find that the quantity \(\Xi^{(p)}_\rho\) given by (\ref{eq:Xipl}) admits a natural interpretation as the expectation value of a Schur polynomial in the unitary matrix model. Explicitly,
\begin{equation}\label{eq:XipSchur}
\Xi^{(p)}_\rho = \frac{1}{\mathcal{Z}(S^3/\mathbb{Z}_p)}
\sum_{\nu}
\mathcal{S}_{0\nu}^2\,
\mathcal{T}_{\nu\nu}^{-p}\,
\frac{\mathcal{S}_{\nu\rho}}{\mathcal{S}_{0\nu}}
=
\big\langle \mathbf{s}_{\rho}(U) \big\rangle_{L(p,1)} .
\end{equation}
Using this relation the normalised $(\alpha,\beta)$ torus knot invariant in $L(p,1)=S^3/\mathbb{Z}_p$ can be written directly in terms of Schur-polynomial expectation values as
\begin{equation}
W^{(\alpha,\beta)}_{\mu}\!\left(S^3/\mathbb{Z}_p\right)
=
q^{(f-\alpha\beta)\kappa_\mu}\sum_{\substack{\rho\\ |\rho|=\alpha|\mu|}}
c^{\rho}_{\mu,\alpha}\;
q^{\frac{\beta}{\alpha}\kappa_{\rho}}\;
\big\langle \mathbf{s}_{\rho}(U) \big\rangle_{L(p,1)}.
\end{equation}
Like (\ref{eq:Xipl}) this result is valid for any \(N\) and \(k\).

\paragraph{\(\bullet\) \(p=1\) case:} Let us first consider the case $p=1$. Using Eqs.~(\ref{eq:dimqdef}) and (\ref{eq:TbyTtoqkappa}), the expectation value of the Schur polynomial can be written as
\begin{eqnarray}\label{eq:schur-dimq-equality}
\big\langle \mathbf{s}_{\rho}(U) \big\rangle_{L(1,1)}
= q^{\kappa_\mu}\,\dim_q \mu .
\end{eqnarray}
This result holds for arbitrary values of $N$ and $k$. It can be checked explicitly by evaluating (\ref{eq:XipSchur}) for $p=1$ and for small values of $N$ and $k$, where the sum over $\nu$ in (\ref{eq:XipSchur}) runs over a finite set of integrable representations. One may also verify this result in the double-scaling limit. In particular, if the representation $\mu$ is small compared to $N$ and $k$ (i.e., the number of boxes in $\mu$ is $\mathcal{O}(1)$), the expectation value $\langle \mathbf{s}_{\rho}(U) \rangle_{L(1,1)}$ can be evaluated using the saddle point solution (\ref{eq:bgsol}). Expanding the right-hand side perturbatively in $N$, one finds that Eq.~(\ref{eq:schur-dimq-equality}) is satisfied order by order in $N$. Thus we get back (\ref{eq:WTST}).

\paragraph{\(\bullet\) \(p>1\) case:} From the structure of the matrix model, we observe that the expectation value $\langle \mathbf{s}_{\rho}(U) \rangle_{L(1,1)}$ depends on the parameter $q = e^{\frac{2\pi i\lambda}{N}}$. Examining the effective action (\ref{eq:Seff}), the corresponding saddle point equation (\ref{eq:sad0}), and its solution (\ref{eq:bgsol}), we find that in the double-scaling limit all correlation functions evaluated on the saddle point depend only on the combination $\lambda/p$. Consequently, expectation values of correlation functions of any gauge-invariant operators for $p>1$ can be obtained from the corresponding $p=1$ results by replacing $\lambda$ with $\lambda/p$\footnote{The last term in $S_{\text{eff}}$ in (\ref{eq:Seff}) does not enter the saddle point equation and therefore does not contribute to any normalized correlation functions.}.

Since $q = e^{\frac{2\pi i\lambda}{N}}$, this replacement amounts to the transformation $q \to q^{1/p}$. Therefore, in the large-$N$ limit we obtain
\begin{equation}
\langle \mathbf{s}_{\rho}(U) \rangle_{L(p,1)}
=
\langle \mathbf{s}_{\rho}(U) \rangle_{L(1,1)}\Big|_{q\to q^{1/p}}
=
q^{\frac{\kappa_\mu}{p}}\,\dim_{q^{1/p}}\mu .
\end{equation}
This relation can be verified explicitly in the double-scaling limit by evaluating both sides of the equation. In particular, one computes the Schur polynomial on the leading saddle point solution (\ref{eq:bgsol}) for small representations $\rho$ \cite{Maji:2023zio}, and compares the result with the corresponding $q$-dimension expanded order by order in the large-$N$ limit. See app. \ref{app:largeNcheck} for some explicit calculations.

Thus, in the large-$N$ limit we can write
\begin{equation}
W^{(\alpha,\beta)}_{\mu}\!\left(S^3/\mathbb{Z}_p,f;q,v\right)
=
q^{(f-\alpha\beta)\kappa_\mu}
\sum_{\substack{\rho\\ |\rho|=\alpha|\mu|}}
c^{\rho}_{\mu,\alpha}\;
\left(q^{\frac{1}{p}}\right)^{\frac{\alpha + p\beta}{\alpha}\kappa_{\rho}}\;
\dim_{q^{\frac{1}{p}}}\rho .
\end{equation}
Using the definition (\ref{eq:torusknotS3}), this expression can be rewritten in terms of torus knot invariants in $S^3$. In particular, we obtain
\begin{eqnarray}\label{eq:WS3modZpandWS3}
W^{(\alpha,\beta)}_{\mu}\!\left(S^3/\mathbb{Z}_p,f;q,v\right)
=
q^{\frac{f(p-1)\kappa_\mu}{p}}
\,q^{\frac{\alpha^2\kappa_\mu}{p}}\,
W^{(\alpha,\alpha+p\beta)}_{\mu}\!\left(S^3;q^{\frac{1}{p}},v^{\frac{1}{p}}\right).
\end{eqnarray}
In the large-$N$ limit the torus knot invariants in the lens space $S^{3}/\mathbb{Z}_{p}$ take a remarkably simple form. In particular, the invariant of an $(\alpha,\beta)$ torus knot in $S^{3}/\mathbb{Z}_{p}$ can be expressed directly in terms of the invariant of the $(\alpha,\alpha+p\beta)$ torus knot in $S^{3}$. This relation establishes a direct correspondence between torus knot sectors in the lens space and those in the three--sphere, with the effect of the $\mathbb{Z}_{p}$ quotient appearing as a shift in the slope of the knot in the large-$N$ regime.

\subsection{Reduced invariants}\label{sec:reducedinvariant}

For the purpose of formulating the knot--quiver correspondence, it is convenient to work with reduced knot invariants \cite{fuji2013super, Kucharski:2017ogk, Kucharski:2025tqb}. We therefore define the reduced invariant $\widetilde{W}^{(\alpha,\beta)}_\mu(S^3/\mathbb{Z}_p;q,v)$ by normalizing the torus knot invariant $W^{(\alpha,\beta)}_{\mu}(S^3/\mathbb{Z}_p,f;q,v)$ with respect to the unknot invariant,
\begin{equation}
\widetilde{W}^{(\alpha,\beta)}_\mu\!\left(S^3/\mathbb{Z}_p; q,v\right)
=
\frac{W^{(\alpha,\beta)}_{\mu}\!\left(S^3/\mathbb{Z}_p,f;q,v\right)}
{W^{(1,0)}_{\mu}\!\left(S^3/\mathbb{Z}_p,f;q,v\right)}.
\end{equation}

Using the relation (\ref{eq:WS3modZpandWS3}), the reduced invariant in the lens space can be expressed in terms of torus knot invariants in $S^3$,
\begin{equation}
\widetilde{W}^{(\alpha,\beta)}_\mu\!\left(S^3/\mathbb{Z}_p; q,v\right)
=
q^{\frac{\alpha^2-1}{p}\kappa_\mu}
\frac{W^{(\alpha,\alpha+p\beta)}_{\mu}\!\left(S^3,f;q^{1/p},v^{1/p}\right)}
{W^{(1,1)}_{\mu}\!\left(S^3,f;q^{1/p},v^{1/p}\right)}.
\end{equation}

Since
\[
W^{(1,1)}_\mu\!\left(S^3,f;q^{1/p},v^{1/p}\right)
=
W^{(1,0)}_\mu\!\left(S^3,f;q^{1/p},v^{1/p}\right),
\]
the expression simplifies to
\begin{equation}
\widetilde{W}^{(\alpha,\beta)}_\mu\!\left(S^3/\mathbb{Z}_p; q,v\right)
=
q^{\frac{\alpha^2-1}{p}\kappa_\mu}
\,
\widetilde{W}^{(\alpha,\alpha+p\beta)}_{\mu}\!\left(S^3;q^{1/p},v^{1/p}\right).
\end{equation}

The reduced invariant $\widetilde{W}$ is independent of the framing $f$. Rescaling the variables $q\to q^p$ and $v\to v^p$, we obtain the key identity
\begin{equation}\label{eq:lensspace-main-identity}
\widetilde{W}^{(\alpha,\beta)}_\mu\!\left(S^3/\mathbb{Z}_p; q^p,v^p\right)
=
q^{(\alpha^2-1)\kappa_\mu}
\,
\widetilde{W}^{(\alpha,\alpha+p\beta)}_{\mu}\!\left(S^3;q,v\right)
\end{equation}
which relates torus knot invariants in the lens space to those in $S^3$.

\section{Quivers for torus knots in lens space}\label{sec:quiver}

The superpolynomial provides a refined extension of the HOMFLY--PT polynomial by encoding the full triply graded homological structure associated with a knot $K$. It is defined as the generating function
\[
\mathcal{P}_K(q,v,t) = \sum_{i,j,k} v^i q^j t^k \, \dim \mathcal{H}^{i,j,k}(K),
\]
where $\mathcal{H}^{i,j,k}(K)$ denotes the triply graded knot homology. In this framework, the HOMFLY--PT polynomial arises as a decategorified limit of the superpolynomial obtained by taking the graded Euler characteristic. Equivalently, one recovers the classical invariant via the specialization \cite{Dunfield:2006, Rasmussen:2010, Gukov:2011ry, Gukov:2015qea}
\[
P_K(q,v) = \mathcal{P}_K(q,v,-1).
\]
Thus, the HOMFLY--PT polynomial captures only the alternating sum of homology dimensions, while the superpolynomial retains the richer graded information of the underlying homological theory.

The superpolynomial for \((2,2n+1)\) torus knot in \(S^3\) is given by
\begin{equation}\label{eq:suppolytorusknot}
	P_{r}^{(2,2 n +1)}(q,v,t) = \frac{a^{2 n r}}{q^{2 n r}} \sum_{0 \leq k_n \leq \dots \leq k_1  \leq r} \left[{r}\atop{k_1}\right]_{q^2} \left[{k_1}\atop{k_2}\right]_{q^2} \cdots \left[{k_{n-1}}\atop{k_n}\right]_{q^2} q^{\mathbf{A}} t^{\mathbf{B}} \prod_{i=1}^{k_1} (1+ a^2 q^{2(i-2)}t)
\end{equation}
where
\begin{equation}
    \mathbf{A} = 2 \sum_{i=1}^{n} \left((2 r +1) k_i - k_{i-1} k_{i}\right), \quad \mathbf{B} = 2 \sum_{i=1}^{n} k_i, \quad \left[{r}\atop{k}\right]_{q^2} = \frac{(q^2;q^2)_r}{(q^2;q^2)_k(q^2;q^2)_{r-k}}.
\end{equation}

The knot invariants obtained from the superpolynomial after setting $t=-1$ are the HOMFLY-PT invariants. They are related to the invariants obtained from the knot operator formalism (\ref{eq:2-2n+1torusknot}) with \(f=0\) by a change of variables:
\begin{equation}\label{eq:changeofvariables}
    q\xrightarrow{}q^{-\frac{1}{2}}~~~\text{and} ~~~v\xrightarrow{}v^{-\frac{1}{2}}.
\end{equation}
More explicitly,
\begin{equation}\label{eq:WPrelation}
    P_{r}^{(2,2 n +1)}(q^{-1/2},v^{-1/2}, -1) = \widetilde W_r^{(2,2n+1)}(S^3,q,v).
\end{equation}
\iffalse
Here we list a few of these invariants.
\begin{itemize}
    \item \(n=1\) : Trefoil
    \begin{center}
{\tiny 
\begin{tabular}{|c|c|}
\hline
$r$ & $P^{(2,2n+1)}_r(S^3)$ \\
\hline
1 & \(\frac{1}{a^2 q}\Bigg[ a-q+aq^2 \Bigg]\)   \\
\hline
2 & \(\frac{1}{a^4 q^4}\Bigg[ q - a (1 + q + q^3 + q^4) + a^2 (1 + q^2 + q^3 + q^6) \Bigg]\) \\
\hline
3 & \(\frac{1}{a^6 q^{10}}\Bigg[ -q + a (1 + q + q^2) (1 + q^4)\) \\
 & \(a^3 q (1 + q^4) (1 + q^2 + q^3 + q^8) - a^2 (1 + q + q^2) (1 + q^3 + q^4 + q^8) \Bigg]\) \\
\hline
\end{tabular}
}
\end{center}
\item \(n=2\) : Cinquefoil
\begin{center}
{\tiny 
\begin{tabular}{|c|c|}
\hline
$r$ & $P^{(2,5)}_r(S^3)$ \\
\hline
 1 & \(\frac{1}{a^3 q^2}\Bigg[ -q (1 + q^2) + a (1 + q^2 + q^4) \Bigg]\) \\
\hline
 2 & \(\frac{1}{a^6 q^8}\Bigg[ q + q^7 - a (1 + q + q^6 + q^7) + a^2 (1 + q^6 + q^{12}) - \) \\
 & \(
 q^2 (1 + q) (-a + q) (-1 + a (1 + q^2 + q^6)) \Bigg]\) \\
\hline
 3 & \(\frac{1}{a^9 q^{19}}\Bigg[ (-q (1 + q^4) (1 + q^2 + q^3 + q^8) + \) \\
& \(
  a^3 (q + q^3 + q^4 + 2 q^5 + q^6 + 2 q^7 + q^8 + 2 q^9 + 2 q^{10} + \) \\
& \(
     2 q^{11} + q^{12} + 2 q^{13} + q^{14} + 2 q^{15} + q^{16} + q^{17} + q^{19} + 
     q^{20} + q^{21} + q^{25}) - \) \\
& \(
  a^2 (1 + q + q^2) (1 + q^4) (1 + 
     q^2 (1 + q + q^3 + q^4 + q^6 + q^8 + q^9 + q^{14})) + \) \\
& \(
  a (1 + q + q^2) (1 + 
     q^2 (1 + q + q^2 + 2 q^4 + q^5 + q^6 + q^8 + q^9 + q^{10} + q^{14}))) \Bigg]\) \\
\hline
\end{tabular}
}
\end{center}
\end{itemize}
These results are in agreement with (\ref{eq:2-2n+1torusknot}) with \(f=0\).
\fi

Before proceed further we point out an important issue here. Let us look at (\ref{eq:2-2n+1torusknot}) and (\ref{eq:suppolytorusknot}) for \(n=1\). Both represent trefoil in symmetric \(r\) representation. Up to some overall factors both the invariants are given by sum over an integer from \(0\) to \(r\). Although the final results match the individual summands do not match. Similar observation persists for high values of \(n\).

\subsection{Quivers}

To determine the quiver structure, we begin by introducing the generating function for torus knot invariants in the lens space $S^3/\mathbb{Z}_p$, obtained by summing the reduced invariants over all symmetric representations,
\begin{equation}
\mathcal{P}(x;q,v)
=
\sum_{r=0}^{\infty}
\frac{\widetilde{W}^{(\alpha,\beta)}_{r}\!\left(S^3/\mathbb{Z}_p; q^p,v^p\right)}{(q^{-1},q^{-1})_r}
x^r 
\end{equation}
where,
\begin{equation}
    (x,y)_r = \prod_{i=0}^{r-1}\left( 1 - x y^i \right).
\end{equation}

Using the relation (\ref{eq:lensspace-main-identity}), the generating function can be rewritten as\footnote{One can still use the relation (\ref{eq:lensspace-main-identity}) for large values of symmetric representations in the double scaling limit \cite{Maji:2023zio}.}
\begin{equation}
\mathcal{P}(x;q,v)
=
\sum_{r=0}^{\infty}
\frac{q^{(\alpha^2-1)\kappa_r}\,
\widetilde{W}^{(\alpha,\alpha+p\beta)}_{r}\!\left(S^3;q,v\right)}
{(q^{-1},q^{-1})_r}
x^r .
\end{equation}
The reduced invariants in $S^3$ can be expressed in terms of HOMFLY--PT polynomials obtained from the superpolynomial (\ref{eq:WPrelation}),
\begin{equation}
\widetilde{W}^{(\alpha,\alpha+p\beta)}_{r}\!\left(S^3;q,v\right)
=
P^{(\alpha,\alpha+p\beta)}_{r}\!\left(q^{-1/2},v^{-1/2},-1\right).
\end{equation}

Substituting this relation into the generating function yields
\begin{equation}
\mathcal{P}(x;q,v)
=
\sum_{r=0}^{\infty}
\frac{q^{(\alpha^2-1)\kappa_r}\,
P^{(\alpha,\alpha+p\beta)}_{r}\!\left(q^{-1/2},v^{-1/2},-1\right)}
{(q^{-1},q^{-1})_r}
x^r .
\end{equation}

Using the knot--quiver correspondence together with the change of variables (\ref{eq:changeofvariables}), the right-hand side can be written in the standard quiver form
\begin{equation}
\mathcal{P}(x;q^{-2},v^{-2})
=
\sum_{d_1,\ldots,d_m\ge0}
x^{\sum_{i=1}^{m} d_i}
q^{\sum_{i,j=1}^{m}(\mathcal{Q}_{ij}-\alpha^2+1)d_i d_j}
\frac{\prod_{i=1}^{m}
q^{(l_i+\alpha^2+1)d_i}
v^{(a_i-\alpha^2+1)d_i}
(-1)^{t_i d_i}}
{\prod_{i=1}^{m}(q^2,q^2)_{d_i}},
\end{equation}
where $\mathcal{Q}_{ij}$ is the quiver matrix associated with the $(\alpha,\alpha+p\beta)$ torus knot in $S^3$.

This relation implies that the quiver structure of an $(\alpha,\beta)$ torus knot in $S^3/\mathbb{Z}_p$ is directly related to the quiver structure of the $(\alpha,\alpha+p\beta)$ torus knot in $S^3$. In particular, the corresponding quiver matrices are related by
\begin{equation}
\mathcal{Q}^{(\alpha,\beta)}_{S^3/\mathbb{Z}_p}
=
\mathcal{Q}^{(\alpha,\alpha+p\beta)}_{S^3}
-
(\alpha^2-1)
\begin{pmatrix}
1 & 1 & \cdots & 1 \\
1 & 1 & \cdots & 1 \\
\vdots & \vdots & \ddots & \vdots \\
1 & 1 & \cdots & 1
\end{pmatrix}.
\end{equation}

The second term corresponds to a universal shift of the quiver matrix and can be removed by an appropriate choice of framing \cite{Kucharski:2017ogk}.

As a simple example, consider the case $\alpha=2$ with $\beta$ odd. For odd values of $p$, the corresponding torus knot invariant in $S^3/\mathbb{Z}_p$ is equivalent to the $(2,2+p\beta)$ torus knot invariant in $S^3$. In $S^3$, such knots are conventionally written as $(2,2n+1)$, with
\begin{equation}
n=\frac{1}{2}(1+p\beta).
\end{equation}

The dimension of the corresponding quiver matrix for these knots is $2n+1 = 2+p\beta$. In particular, for the trefoil knot in $S^3/\mathbb{Z}_p$ (corresponding to $\beta=3$), the dimension of the associated quiver matrix is $2+3p$ for odd values of $p$.

\section{Conclusion}\label{sec:conclusion}

In this work we have investigated torus knot invariants in the lens space $S^3/\mathbb{Z}_p$ within the framework of Chern--Simons theory. Using the surgery description of lens spaces and the modular properties of the torus Hilbert space, we first derived a general expression for the invariant of an $(\alpha,\beta)$ torus knot in this background. This formula is exact and valid for arbitrary values of the rank $N$ and level $k$ of the Chern--Simons theory, although its direct evaluation involves summations over integrable representations of the affine algebra.

A substantial simplification arises in the double-scaling limit (\ref{eq:DSlimit}). In this regime the modular sums appearing in the lens space invariants admit a natural description in terms of a unitary matrix model. Using this formulation, we showed that the correlation functions entering the knot invariants depend only on the combination $\lambda/p$. Consequently, the expectation values relevant for torus knot invariants in $S^3/\mathbb{Z}_p$ can be obtained from the corresponding $p=1$ results through the simple replacement $q \to q^{1/p}$.

This observation leads to a remarkably simple large-$N$ relation between torus knot invariants in the lens space and those in the three--sphere. In particular, we find that the invariant of an $(\alpha,\beta)$ torus knot in $S^3/\mathbb{Z}_p$ can be expressed directly in terms of the invariant of the $(\alpha,\alpha+p\beta)$ torus knot in $S^3$. Thus, in the large-$N$ limit the effect of the $\mathbb{Z}_p$ quotient manifests itself as a simple shift in the slope of the torus knot, providing a direct map between torus knot sectors in the lens space and those in $S^3$. 

Motivated by the knot--quiver correspondence, we then introduced reduced invariants obtained by normalizing the torus knot invariant by the corresponding unknot invariant. In terms of these reduced invariants we derived a compact identity relating torus knot invariants in $S^3/\mathbb{Z}_p$ to those in $S^3$. This relation allows the generating function of lens space torus knot invariants to be expressed in the standard form of a quiver partition function after an appropriate redefinition of variables.

Using this structure, we determine the quiver associated with torus knots in $S^3/\mathbb{Z}_p$. In particular, we show that the quiver matrix for an $(\alpha,\beta)$ torus knot in the lens space is directly related to the quiver matrix for the $(\alpha,\alpha+p\beta)$ torus knot in $S^3$, up to a universal shift that can be absorbed into a framing redefinition. Since the quiver structure is independent of the rank $N$ and level $k$ of Chern--Simons theory, this relation allows us to determine the quiver data for torus knots in lens spaces from the double-scaling limit.

Our results therefore provide a natural extension of the knot--quiver correspondence to lens space backgrounds. More broadly, they reveal a simple and universal relation between torus knot invariants in $S^3/\mathbb{Z}_p$ and those in $S^3$, suggesting that many structural properties of knot invariants in nontrivial three--manifolds may ultimately be understood in terms of their counterparts in the three--sphere.

Several directions merit further investigation. It would be interesting to extend the present analysis to more general knots and links in lens spaces and other Seifert manifolds and the corresponding knot complements \cite{Ekholm:2021irc, Kucharski:2020rsp, Ekholm:2020lqy}. Another natural problem is to understand the role of the large-$N$ phase structure of the Chern--Simons matrix model and its implications for knot invariants and their quiver descriptions. Finally, it would be worthwhile to explore whether the relation uncovered here admits a deeper interpretation in the context of large-$N$ dualities and topological string theory.

\paragraph{Acknowledgment:} SD thanks Piotr Su{\l}kowski for fruitful discussions on the knot--quiver correspondence during his visit to Warsaw in 2025 and acknowledges the hospitality of the University of Warsaw during this visit. The work of RB is supported by the SERB project SERB/PHY/2023-2024/65. We thank Dheeraj Kulkarni for useful discussions. We also acknowledge the use of AI tools (ChatGPT and Grammarly) solely for improving the grammar and clarity of the manuscript. Finally, we are grateful to the people of India for their continued support of research in basic science. 

\appendix

\section{Construction of $S^3/\mathbb{Z}_p$}\label{app:lens-spac}

Let $D^{2}\times S^{1}$ denote a solid torus whose boundary is the two--torus
\[
\partial(D^{2}\times S^{1})=\mathbb{T}^{2}.
\]
Loops on the torus are classified by the first homology group
\[
H_{1}(\mathbb{T}^{2})=\mathbb{Z}\times\mathbb{Z}.
\]
A basis for this homology group is given by the \emph{meridian} $m$ and the \emph{longitude} $l$. For a solid torus, the meridian is the contractible cycle bounding a disk in $D^{2}\times S^{1}$, while the longitude is the non--contractible cycle that bounds a punctured disk in the solid torus.

An $(\alpha,\beta)$ torus knot is defined as the isotopy class of an embedded simple closed curve $\gamma$ in the solid torus whose projection to the boundary torus represents the homology class
\[
\llbracket\gamma\rrbracket = \alpha \llbracket l \rrbracket + \beta \llbracket m\rrbracket,
\]
with $\gcd(\alpha,\beta)=1$. Such a knot winds $\alpha$ times along the longitudinal direction and $\beta$ times around the meridional direction of the torus.

In his seminal work \cite{Witten:1988hf}, Witten exploited the fact that the three--sphere $S^{3}$ admits a genus--one Heegaard splitting. In this description, $S^{3}$ is obtained by gluing two solid tori $(D^{2}\times S^{1})_{L}$ and $(D^{2}\times S^{1})_{R}$ along their common boundary torus $\mathbb{T}^{2}$ through an orientation--reversing diffeomorphism. This perspective plays a central role in the operator formulation of Chern--Simons theory, where the gluing map is interpreted as a modular transformation acting on the Hilbert space $\mathcal{H}(\mathbb{T}^{2})$ associated with the boundary torus.

Let us choose meridian--longitude bases
\[
(\llbracket m_L \rrbracket,\llbracket l_L \rrbracket), 
\qquad
(\llbracket m_R \rrbracket,\llbracket l_R \rrbracket)
\]
for the homology cycles of the boundary tori of the solid tori
$(D^{2}\times S^{1})_{L}$ and $(D^{2}\times S^{1})_{R}$, respectively.

\paragraph{Lens space:}The lens space $L(p,1)$ is the three--manifold obtained by gluing the two solid tori such that the meridian of the left torus is mapped to the cycle \cite{Rolfsen2003KnotsandLinks}
\[
\llbracket m_L \rrbracket \;\mapsto\; 
\llbracket m_R \rrbracket + p\,\llbracket l_R \rrbracket .
\]
With respect to the chosen bases, this condition specifies that the meridian of the left torus wraps once around the meridian and $p$ times around the longitude of the right torus. This distinction is essential: the meridian bounds a disk inside the solid torus, while the longitude represents the non--contractible cycle.

The above prescription determines the first column of the corresponding mapping class group element in $SL(2,\mathbb{Z})$. A convenient representative of this element is
\[
\begin{pmatrix}
1 & p \\
0 & 1
\end{pmatrix}
\in SL(2,\mathbb{Z}),
\]
which acts on the homology cycles
\[
\begin{pmatrix}
\llbracket m \rrbracket \\
\llbracket l \rrbracket
\end{pmatrix}
\quad \longmapsto \quad
\begin{pmatrix}
1 & p \\
0 & 1
\end{pmatrix}
\begin{pmatrix}
\llbracket m \rrbracket \\
\llbracket l \rrbracket
\end{pmatrix},
\]
so that $\llbracket m \rrbracket \mapsto \llbracket m \rrbracket + p\,\llbracket l \rrbracket$.

Introducing the standard generators of $SL(2,\mathbb{Z})$,
\[
S=
\begin{pmatrix}
0 & 1\\
-1 & 0
\end{pmatrix},
\qquad
T=
\begin{pmatrix}
1 & 0\\
1 & 1
\end{pmatrix},
\]
the matrix $S$ exchanges the $(1,0)$ and $(0,1)$ cycles of the torus, while $T$ acts as a Dehn twist that shifts the meridian by the longitude. One can readily verify that
\[
(TST)^{p}
=
\begin{pmatrix}
1 & p\\
0 & 1
\end{pmatrix}.
\]
Thus, up to isotopy, the lens space
\[
L(p,1)\cong S^{3}/\mathbb{Z}_{p}
\]
can be obtained by gluing the left and right solid tori using the above mapping class group element.

In Chern--Simons theory, diffeomorphisms of the boundary torus are represented by operators acting on the Hilbert space $\mathcal{H}(\mathbb{T}^{2})$. In particular,
\[
\mathcal{S} \equiv U_S , \qquad
\mathcal{T} \equiv U_T ,
\]
denote the operators implementing the modular transformations $S$ and $T$ on the Hilbert space and is given by (\ref{eq:Sdef}) and (\ref{eq:Tdef}) respectively. These operators realize the geometric actions of inversion and Dehn twist on states in the torus Hilbert space and play a central role in the operator formulation of knot invariants.

\section{To check \(\langle{\mathbf{s}_\rho(U)}\rangle_{L(p,1)} = \left(q^{\kappa_\rho} \dim_q(\rho)\right)\big\vert_{q \to q^{1/p}}\) in the large \(N\) limit}\label{app:largeNcheck}

We have derived the result 
\begin{equation}
    \langle\mathbf{s}_\rho(U)\rangle_{L(p,1)} = \left(q^{\kappa_\rho} \dim_q(\rho)\right)\big\vert_{q \to q^{1/p}}
\end{equation}
The LHS can be computed upto leading and sub-leading orders in $N$ using saddle point analysis and using the following character expansion of schur polynomials  
\[
        \mathbf{s}_{\rho}(U) = \sum_{\vec{k}} \frac{
            \chi_\rho(c(\vec{k}))
        }{
            \prod_{i} k_i! \; i^{k_i}
        } \prod_{j} (\mathrm{Tr}(U^j)^{k_j}
\]
where the summation vector $\vec{k}$ is such that $\sum_r r k_r = |\rho|$. Using this we find that
\begin{equation}
    \begin{gathered}
        \mathbf{s}_{\yng(1)}(U) = \mathrm{Tr}(U) \\
        \mathbf{s}_{\yng(2)}(U) = \frac{\mathrm{Tr}(U)^2}{2} + \frac{\mathrm{Tr}(U^2)}{2}\\
        \mathbf{s}_{\yng(1,1)}(U) = \frac{\mathrm{Tr}(U)^2}{2} - \frac{\mathrm{Tr}(U^2)}{2}
    \end{gathered}
\end{equation}
In large-$N$ limit of our matrix model admits the following expansion for any observable $O$,
\[
        \langle O \rangle = \langle O \rangle_{0} + \frac{\langle O\rangle_{1}}{N
        ^2} + \frac{\langle O \rangle_{2}}{N^4} + \cdots 
\]
Further in order to remove $N$ dependece from the expectation of traces we define a modified trace
\[
            \mathrm{tr}(U^m) = \frac{1}{N} \mathrm{Tr}(U^m)
\]
Thus we have
\begin{equation} \label{eq:schurfund}
    \begin{aligned}
        \frac{1}{N} \langle \mathbf{s}_{\yng(1)}(U) \rangle 
        &=  \langle \mathrm{tr}(U) \rangle \\
        &\approx \langle \mathrm{tr}(U)\rangle_0 + 
        \frac{\langle \mathrm{tr}(U) \rangle_1}{N^2} + \cdots
    \end{aligned}
\end{equation}
\begin{equation} \label{eq:schurbbox}
    \begin{aligned}
        \frac{2}{N^2} \langle\mathbf{s}_{\yng(2)}(U)\rangle 
        &= \langle \mathrm{tr}(U)^2 \rangle 
        + \frac{1}{N} \langle \mathrm{tr}(U^2)\rangle\\
        &= \left\langle \mathrm{tr}(U) \right\rangle^2
            + \frac{\langle\mathrm{tr} U \mathrm{tr} U \rangle_c}{N^2}
            + \frac{1}{N} \left\langle \mathrm{tr}(U^2) \right\rangle \\
        &\approx
        \left(
        \left\langle \mathrm{tr} U \right\rangle_0^2
        + \frac{2}{N^2}
          \left\langle \mathrm{tr} U \right\rangle_0
         \left\langle \mathrm{tr} U \right\rangle_1
        + \cdots
        \right)
        + \frac{\left\langle \mathrm{tr} U \, \mathrm{tr} U \right\rangle_{c,0}}{N^2}\\
        & \qquad \qquad + \frac{1}{N} \left(
            \langle\mathrm{tr}(U^2)\rangle_0 + \frac{\langle\mathrm{tr}(U^2)\rangle_1}{N^2} + \cdots
        \right)\\
        &= \langle\mathrm{tr} U\rangle^2_0 
        + \frac{\langle\mathrm{tr}(U^2)\rangle_0}{N} 
        + \frac{
        \left[ 
        2 \langle\mathrm{tr} U\rangle_0 \langle\mathrm{tr} U\rangle_1 
        + \langle\mathrm{tr} U \mathrm{tr} U\rangle_{c,0}\right]
        }{N^2} 
        + \frac{\langle\mathrm{tr}(U^2)\rangle_1}{N^3}
        + \mathcal{O}(\frac{1}{N^4})
    \end{aligned}
\end{equation}
\begin{equation} \label{eq:schurboxbox}
   \begin{aligned}
       \frac{2}{N^2} \langle\mathbf{s}_{\yng(1,1)}(U)\rangle 
       &= \langle\mathrm{tr}(U)^2\rangle - \frac{1}{N} \langle\mathrm{tr}(U^2)\rangle \\
       &= \langle\mathrm{tr} U\rangle^2 
        + \frac{\langle\mathrm{tr} U \mathrm{tr} U\rangle_c}{N^2} 
        - \frac{\langle\mathrm{tr}(U^2)\rangle}{N} \\
        &\approx 
        \left(
            \langle\mathrm{tr} U\rangle^2_0 
            + \frac{2}{N^2} \langle\mathrm{tr} U\rangle_0 \langle\mathrm{tr} U\rangle_1 
            + \cdots
        \right)
        + \frac{\langle\mathrm{tr} U \mathrm{tr} U\rangle_{c,0}}{N^2} \\
        & \qquad \qquad - \frac{1}{N} \left(
            \langle\mathrm{tr}(U^2)\rangle_0 + \frac{\langle\mathrm{tr}(U^2)\rangle_1}{N^2} + \cdots
        \right)\\
        &= \langle\mathrm{tr} U\rangle^2_0 
        -\frac{\langle\mathrm{tr}(U^2)\rangle_0 }{N} 
        + \frac{
        \left[ 
        2  \langle\mathrm{tr} U\rangle_0 \langle\mathrm{tr} U\rangle_1 
        + \langle\mathrm{tr} U \mathrm{tr} U\rangle_{c,0}
        \right]
        }{N^2} 
        - \frac{\langle\mathrm{tr}(U^2)\rangle_1}{N^3}
        + \mathcal{O}(\frac{1}{N^4})\\
   \end{aligned}
\end{equation}
It has been shown in \cite{Maji:2023zio} that
\begin{equation}
    \begin{gathered}
    \langle\mathrm{tr} U\rangle_0 = \frac{
            i \left(
            1 - e^{\frac{2  \pi i  \lambda}{p}} 
            \right)
            }{2 \pi  \lambda }  p  \\
    \langle\mathrm{tr}(U^2)\rangle_0 = \frac{
    i  \left(
        4 e^{\frac{2  \pi i \lambda }{p}} 
        - 3 e^{\frac{4  \pi i  \lambda }{p}}
        - 1
    \right)
    }{4 \pi  \lambda }  p \\
    \langle\mathrm{tr} U\rangle_1 = \frac{i 
    \left(
        1 - e^{\frac{2 \pi i  \lambda }{p}}
    \right) \pi \lambda  
    }{12 p} \\
    \langle\mathrm{tr} U \mathrm{tr} U\rangle_{c,0} 
    =   e^{\frac{2 \pi i \lambda}{p}} \left(e^{\frac{2 \pi i \lambda}{p}} - 1 \right)
    \end{gathered}
\end{equation}
Plugging these back in (\ref{eq:schurfund}), (\ref{eq:schurbbox}) and (\ref{eq:schurboxbox}), we have
\begin{equation}
    \langle\mathbf{s}_{\yng(1)}(U)\rangle_{L(p,1)} 
    = -\frac{ \pi i  \lambda  \left(-1+e^{\frac{2  \pi  i \lambda }{p}}\right)}{12 N p}
    -\frac{i N p \left(-1+e^{\frac{2  \pi i  \lambda }{p}}\right)}{2 \pi  \lambda }
\end{equation}
\begin{equation}
    \langle\mathbf{s}_{\yng(2)}(U)\rangle_{L(p,1)} 
    = \begin{aligned}[t]
        &-\frac{N^2 p^2 
        \left(-1+e^{\frac{2 \pi i  \lambda }{p}}\right)^2
        }{8 \pi^2 \lambda^2} 
        -\frac{i N p 
        \left(-4 e^{\frac{2  \pi i  \lambda }{p}}
        +3 e^{\frac{4  \pi i  \lambda }{p}}
        +1\right)
        }{8 \pi  \lambda } \\
        &+ \frac{\pi i  \lambda  
        \left(-2 e^{\frac{2  \pi i  \lambda }{p}}
        +3 e^{\frac{4 \pi i  \lambda }{p}}
        -1\right)
        }{12 N p}
        -\frac{
        \pi^2 \lambda^2 
        \left(-1+e^{\frac{2 \pi i \lambda }{p}}\right)^2
        }{288 N^2 p^2}
        \\
        &+ \frac{1}{24} 
        \left(
        -10 e^{\frac{2  \pi i  \lambda }{p}} 
        + 11 e^{\frac{4  \pi i  \lambda }{p}} - 1
        \right)
    \end{aligned}
\end{equation}
\begin{equation}
    \langle\mathbf{s}_{\yng(1,1)}(U)\rangle_{L(p,1)} 
    = \begin{aligned}[t]
        &-\frac{N^2 p^2 
        \left(-1+e^{\frac{2  \pi i  \lambda }{p}}\right)^2
        }{8\pi^2 \lambda^2}
        +\frac{i N p 
        \left(-4 e^{\frac{2  \pi i  \lambda }{p}}
        +3 e^{\frac{4  \pi i  \lambda }{p}}
        + 1 \right)
        }{8 \pi  \lambda }\\
        &-\frac{\pi i  \lambda  
        \left(-2 e^{\frac{2  \pi i  \lambda }{p}}
        +3 e^{\frac{4  \pi i  \lambda }{p}}
        -1\right)
        }{12 N p}
        -\frac{\pi^2 \lambda^2 
        \left(
        -1+e^{\frac{2  \pi i  \lambda }{p}}
        \right)^2
        }{288 N^2 p^2} \\
        &+\frac{1}{24} 
        \left(-10 e^{\frac{2  \pi i  \lambda }{p}}
        +11 e^{\frac{4  \pi i  \lambda }{p}}
        -1 \right)        
    \end{aligned}
\end{equation}
Using (\ref{eq:kappadef}) and (\ref{eq:dimqdef}) one can calculate \(\left(q^{\kappa_\rho} \dim_q(\rho)\right)\big\vert_{q \to q^{1/p}}\) order by order in \(N\) for \(\rho = \yng(1), ~ \yng(2), ~ \yng(1,1)\) and check that the expressions match exactly.

\bibliographystyle{hieeetr}
\bibliography{KQCReferencesv1}

@inproceedings{Kucharski:2025tqb,
	author = "Kucharski, Piotr and Noshchenko, Dmitry",
	title = "{Knot-quiver correspondence: a brief review}",
	eprint = "2505.05668",
	archivePrefix = "arXiv",
	primaryClass = "hep-th",
	month = "5",
	year = "2025"
}

@article{Kucharski:2020rsp,
    author = "Kucharski, Piotr",
    title = "{Quivers for 3-manifolds: the correspondence, BPS states, and 3d $ \mathcal{N} $ = 2 theories}",
    eprint = "2005.13394",
    archivePrefix = "arXiv",
    primaryClass = "hep-th",
    doi = "10.1007/JHEP09(2020)075",
    journal = "JHEP",
    volume = "09",
    pages = "075",
    year = "2020"
}

@article{Ekholm:2021irc,
    author = "Ekholm, Tobias and Gruen, Angus and Gukov, Sergei and Kucharski, Piotr and Park, Sunghyuk and Sto{\v{s}}i{\'c}, Marko and Su{\l}kowski, Piotr",
    title = "{Branches, quivers, and ideals for knot complements}",
    eprint = "2110.13768",
    archivePrefix = "arXiv",
    primaryClass = "hep-th",
    doi = "10.1016/j.geomphys.2022.104520",
    journal = "J. Geom. Phys.",
    volume = "177",
    pages = "104520",
    year = "2022"
}

@article{Ekholm:2020lqy,
    author = "Ekholm, Tobias and Gruen, Angus and Gukov, Sergei and Kucharski, Piotr and Park, Sunghyuk and Su{\l}kowski, Piotr",
    title = "{${\widehat{Z}}$ at Large N: From Curve Counts to Quantum Modularity}",
    eprint = "2005.13349",
    archivePrefix = "arXiv",
    primaryClass = "hep-th",
    doi = "10.1007/s00220-022-04469-9",
    journal = "Commun. Math. Phys.",
    volume = "396",
    number = "1",
    pages = "143--186",
    year = "2022"
}

@Article{Ekholm2020,
	author={Ekholm, Tobias
	and Kucharski, Piotr
	and Longhi, Pietro},
	title={Physics and Geometry of Knots-Quivers Correspondence},
	journal={Communications in Mathematical Physics},
	year={2020},
	month={Oct},
	day={01},
	volume={379},
	number={2},
	pages={361-415},
	issn={1432-0916},
	doi={10.1007/s00220-020-03840-y},
	url={https://doi.org/10.1007/s00220-020-03840-y}
}

@article{Aganagic:2002mv,
  author         = "Aganagic, Mina and Klemm, Albrecht and Marino, Marcos and Vafa, Cumrun",
  title          = "{Matrix model as a mirror of Chern-Simons theory}",
  eprint         = "hep-th/0211098",
  archivePrefix  = "arXiv",
  doi            = "10.1088/1126-6708/2004/02/010",
  journal        = "JHEP",
  volume         = "02",
  pages          = "010",
  year           = "2004"
}

@article{Halmagyi:2003qt,
  author         = "Halmagyi, Nick and Okuda, Takuya and Yasnov, Vadim",
  title          = "{Large-$N$ duality, lens spaces and the Chern-Simons matrix model}",
  eprint         = "hep-th/0312145",
  archivePrefix  = "arXiv",
  doi            = "10.1088/1126-6708/2004/04/014",
  journal        = "JHEP",
  volume         = "04",
  pages          = "014",
  year           = "2004"
}

@article{Brini:2011wi,
    author = "Brini, Andrea and Eynard, Bertrand and Marino, Marcos",
    title = "{Torus knots and mirror symmetry}",
    eprint = "1105.2012",
    archivePrefix = "arXiv",
    primaryClass = "hep-th",
    doi = "10.1007/s00023-012-0171-2",
    journal = "Annales Henri Poincare",
    volume = "13",
    pages = "1873--1910",
    year = "2012"
}

@article{Rosso:1993vn,
    author = "Rosso, Marc and Jones, Vaughan",
    title = "{On the invariants of torus knots derived from quantum groups}",
    doi = "10.1142/S0218216593000064",
    journal = "J. Knot Theor. Ramifications",
    volume = "2",
    pages = "97",
    year = "1993"
}

@article{kauffman1990invariant,
  title={An invariant of regular isotopy},
  author={Kauffman, Louis H},
  journal={Transactions of the American Mathematical Society},
  volume={318},
  number={2},
  pages={417--471},
  year={1990}
}

@article{alexander1928topological,
  title={Topological invariants of knots and links},
  author={Alexander, James W},
  journal={Transactions of the American Mathematical Society},
  volume={30},
  number={2},
  pages={275--306},
  year={1928}
}

@article{jones1985polynomial,
    author = "Jones, V. F. R.",
    title = "{A polynomial invariant for knots via von Neumann algebras}",
    doi = "10.1090/S0273-0979-1985-15304-2",
    journal = "Bull. Am. Math. Soc.",
    volume = "12",
    pages = "103--111",
    year = "1985"
}

@article{marino2001framed,
  title={Framed knots at large N},
  author={Marino, Marcos and Vafa, Cumrun},
  journal={arXiv preprint hep-th/0108064},
  year={2001}
}

@article{RamaDevi:1992np,
    author = "Rama Devi, P. and Govindarajan, T. R. and Kaul, R. K.",
    title = "{Three-dimensional Chern-Simons theory as a theory of knots and links. 3. Compact semisimple group}",
    eprint = "hep-th/9212110",
    archivePrefix = "arXiv",
    reportNumber = "IMSC-92-55",
    doi = "10.1016/0550-3213(93)90652-6",
    journal = "Nucl. Phys. B",
    volume = "402",
    pages = "548--566",
    year = "1993"
}

@article{labastida2001knots,
  title={Knots, links and branes at large N},
  author={Labastida, Jos{\'e} MF and Marino, Marcos and Vafa, Cumrun},
  journal={Journal of High Energy Physics},
  volume={2000},
  number={11},
  pages={007},
  year={2001},
  publisher={IOP Publishing}
}

@article{Ramadevi:1994zb,
    author = "Ramadevi, P. and Govindarajan, T. R. and Kaul, R. K.",
    title = "{Representations of composite braids and invariants for mutant knots and links in Chern-Simons field theories}",
    eprint = "hep-th/9412084",
    archivePrefix = "arXiv",
    reportNumber = "IMSC-94-45",
    doi = "10.1142/S0217732395001769",
    journal = "Mod. Phys. Lett. A",
    volume = "10",
    pages = "1635--1658",
    year = "1995"
}

@article{Nawata:2013qpa,
    author = "Nawata, Satoshi and Ramadevi, P. and Zodinmawia",
    title = "{Colored HOMFLY polynomials from Chern-Simons theory}",
    eprint = "1302.5144",
    archivePrefix = "arXiv",
    primaryClass = "hep-th",
    reportNumber = "NIKHEF-2013-006",
    doi = "10.1142/S0218216513500788",
    journal = "J. Knot Theor.",
    volume = "22",
    pages = "1350078",
    year = "2013"
}

@article{Kucharski:2017ogk,
    author = "Kucharski, Piotr and Reineke, Markus and Stosic, Marko and Su{\l}kowski, Piotr",
    title = "{Knots-quivers correspondence}",
    eprint = "1707.04017",
    archivePrefix = "arXiv",
    primaryClass = "hep-th",
    reportNumber = "CALT-2017-040",
    doi = "10.4310/ATMP.2019.v23.n7.a4",
    journal = "Adv. Theor. Math. Phys.",
    volume = "23",
    number = "7",
    pages = "1849--1902",
    year = "2019"
}

@article{fuji2013super,
	title={Super-A-polynomial for knots and BPS states},
	author={Fuji, Hiroyuki and Gukov, Sergei and Su{\l}kowski, Piotr},
	journal={Nuclear Physics B},
	volume={867},
	number={2},
	pages={506--546},
	year={2013},
	publisher={Elsevier}
}

@article{Maji:2023zio,
	title = {$U(N)$ torus link invariants in the large N limit from the matrix model approach},
	author = {Maji, Archana and Chakraborty, Kushal and Dutta, Suvankar and Ramadevi, P.},
	journal = {Phys. Rev. D},
	volume = {109},
	issue = {6},
	pages = {065021},
	numpages = {22},
	year = {2024},
	month = {Mar},
	publisher = {American Physical Society},
	doi = {10.1103/PhysRevD.109.065021},
	url = {https://link.aps.org/doi/10.1103/PhysRevD.109.065021}
}

@article{Witten:1988hf,
	author = "Witten, Edward",
	editor = "Mitra, Asoke N.",
	title = "{Quantum Field Theory and the Jones Polynomial}",
	reportNumber = "IASSNS-HEP-88-33",
	doi = "10.1007/BF01217730",
	journal = "Commun. Math. Phys.",
	volume = "121",
	pages = "351--399",
	year = "1989"
}

@article{Stevan:2013tha,
	author = "Stevan, S{\'e}bastien",
	title = "{Torus Knots in Lens Spaces and Topological Strings}",
	eprint = "1308.5509",
	archivePrefix = "arXiv",
	primaryClass = "hep-th",
	doi = "10.1007/s00023-014-0362-0",
	journal = "Annales Henri Poincare",
	volume = "16",
	number = "8",
	pages = "1937--1967",
	year = "2015"
}

@article{Labastida:1990bt,
    author = "Labastida, J. M. F. and Llatas, P. M. and Ramallo, A. V.",
    title = "{Knot operators in Chern-Simons gauge theory}",
    reportNumber = "CERN-TH-5756/90, IEM-FT-20/90, US-FT-7/90",
    doi = "10.1016/0550-3213(91)90209-G",
    journal = "Nucl. Phys. B",
    volume = "348",
    pages = "651--692",
    year = "1991"
}

@article{Kaul:1991np,
    author = "Kaul, R. K. and Govindarajan, T. R.",
    title = "{Three-dimensional Chern-Simons theory as a theory of knots and links}",
    eprint = "hep-th/9111063",
    archivePrefix = "arXiv",
    reportNumber = "IMSC-91-33, IISC-CTS-7-91",
    doi = "10.1016/0550-3213(92)90524-F",
    journal = "Nucl. Phys. B",
    volume = "380",
    pages = "293--333",
    year = "1992"
}

@article{Naculich:2007nc,
    author = "Naculich, Stephen G. and Schnitzer, Howard J.",
    title = "{Level-rank duality of the U(N) WZW model, Chern-Simons theory, and 2-D qYM theory}",
    eprint = "hep-th/0703089",
    archivePrefix = "arXiv",
    reportNumber = "BRX-TH-582, BOW-PH-139",
    doi = "10.1088/1126-6708/2007/06/023",
    journal = "JHEP",
    volume = "06",
    pages = "023",
    year = "2007"
}

@article{Stevan:2010jh,
    author = "Stevan, Sebastien",
    title = "{Chern-Simons Invariants of Torus Links}",
    eprint = "1003.2861",
    archivePrefix = "arXiv",
    primaryClass = "hep-th",
    doi = "10.1007/s00023-010-0058-z",
    journal = "Annales Henri Poincare",
    volume = "11",
    pages = "1201--1224",
    year = "2010"
}

@phdthesis{Stevan:2014xma,
    author = "Stevan, S{\'e}bastien",
    title = "{Knot invariants, Chern{\textendash}Simons theory and the topological recursion}",
    school = "Geneva U.",
    year = "2014"
}

@article{Naculich:1990pa,
    author = "Naculich, Stephen G. and Riggs, H. A. and Schnitzer, H. J.",
    title = "{Group Level Duality in {WZW} Models and {Chern-Simons} Theory}",
    reportNumber = "BRX-TH-293",
    doi = "10.1016/0370-2693(90)90623-E",
    journal = "Phys. Lett. B",
    volume = "246",
    pages = "417--422",
    year = "1990"
}

@misc{Stokman:1990pa,
    author = "Stokman, Jasper",
    title = "{LECTURE NOTES ON QUIVER REPRESENTATIONS}",
    year = "2021"
}

@article{MLAWER1991863,
author = {Eli J. Mlawer and Stephen G. Naculich and Harold A. Riggs and Howard J. Schnitzer},
title = {Group-level duality of WZW fusion coefficients and Chern-Simons link observables},
journal = {Nuclear Physics B},
volume = {352},
number = {3},
pages = {863-896},
year = {1991},
issn = {0550-3213},
doi = {https://doi.org/10.1016/0550-3213(91)90110-J}
}

@article{Liu:2007kv,
    author = "Liu, Kefeng and Peng, Pan",
    title = "{Proof of the Labastida-Mari{\~n}o-Ooguri-Vafa conjecture}",
    eprint = "0704.1526",
    archivePrefix = "arXiv",
    primaryClass = "math.QA",
    journal = "J. Diff. Geom.",
    volume = "85",
    number = "3",
    pages = "479--525",
    year = "2010"
}

@article{Reshetikhin:1991tc,
    author = "Reshetikhin, N. and Turaev, V. G.",
    title = "{Invariants of three manifolds via link polynomials and quantum groups}",
    doi = "10.1007/BF01239527",
    journal = "Invent. Math.",
    volume = "103",
    pages = "547--597",
    year = "1991"
}

@article{Rozansky_1996,
   title={A contribution of the trivial connection to the Jones polynomial and Witten’s invariant of 3d manifolds, I},
   volume={175},
   ISSN={1432-0916},
   url={http://dx.doi.org/10.1007/BF02102409},
   DOI={10.1007/bf02102409},
   number={2},
   journal={Communications in Mathematical Physics},
   publisher={Springer Science and Business Media LLC},
   author={Rozansky, L.},
   year={1996},
   month=jan, pages={275–296} }

@article{Chakraborty:2021oyq,
    author = "Chakraborty, Kushal and Dutta, Suvankar",
    title = "{New phase in Chern-Simons theory on lens space}",
    eprint = "2102.11088",
    archivePrefix = "arXiv",
    primaryClass = "hep-th",
    doi = "10.1103/PhysRevD.104.026010",
    journal = "Phys. Rev. D",
    volume = "104",
    number = "2",
    pages = "026010",
    year = "2021"
}

@article{Chakraborty:2021bxe,
    author = "Chakraborty, Kushal and Dutta, Suvankar",
    title = "{Large N correlators of Chern-Simons theory in lens spaces}",
    eprint = "2111.11803",
    archivePrefix = "arXiv",
    primaryClass = "hep-th",
    doi = "10.1103/PhysRevD.106.025009",
    journal = "Phys. Rev. D",
    volume = "106",
    number = "2",
    pages = "025009",
    year = "2022"
}

@article{Blau:2016vfc,
    author = "Blau, Matthias and Thompson, George",
    title = "{Chern-Simons Theory with Complex Gauge Group on Seifert Fibred 3-Manifolds}",
    eprint = "1603.01149",
    archivePrefix = "arXiv",
    primaryClass = "hep-th",
    month = "3",
    year = "2016"
}

@article{Blau:2013oha,
    author = "Blau, Matthias and Thompson, George",
    title = "{Chern-Simons Theory on Seifert 3-Manifolds}",
    eprint = "1306.3381",
    archivePrefix = "arXiv",
    primaryClass = "hep-th",
    doi = "10.1007/JHEP09(2013)033",
    journal = "JHEP",
    volume = "09",
    pages = "033",
    year = "2013"
}

@book{Rolfsen2003KnotsandLinks,
  author    = {Dale Rolfsen},
  title     = {Knots and Links},
  publisher = {American Mathematical Society},
  address   = {Providence, RI},
  year      = {2003},
  series    = {AMS Chelsea Publishing},
  isbn      = {978-0-8218-3436-7}
}

@article{Dunfield:2006,
  author = {Dunfield, Nathan M. and Gukov, Sergei and Rasmussen, Jacob},
  title = {The Superpolynomial for Knot Homologies},
  journal = {Experimental Mathematics},
  volume = {15},
  number = {2},
  pages = {129--159},
  year = {2006},
  doi = {10.1080/10586458.2006.10128956},
  eprint = {math/0505662},
  archivePrefix = {arXiv},
  primaryClass = {math.GT}
}

@article{Rasmussen:2010,
  author = {Rasmussen, Jacob},
  title = {Some differentials on Khovanov--Rozansky homology},
  journal = {Geometry \& Topology},
  volume = {19},
  number = {6},
  pages = {3031--3104},
  year = {2015},
  doi = {10.2140/gt.2015.19.3031},
  eprint = {math/0607544},
  archivePrefix = {arXiv},
  primaryClass = {math.GT}
}

@article{Gukov:2011ry,
  author = {Gukov, Sergei and Stosic, Marko},
  title = {Homological Algebra of Knots and BPS States},
  booktitle = {Proceedings of Symposia in Pure Mathematics},
  volume = {85},
  pages = {125--172},
  year = {2012},
  publisher = {American Mathematical Society},
  eprint = {1112.0030},
  archivePrefix = {arXiv},
  primaryClass = {math.QA},
  doi = {10.1090/pspum/085/1397902}
}

@article{Gukov:2015qea,
  author = {Gukov, Sergei and Nawata, Satoshi and Saberi, Iman and Stosic, Marko and Sukowski, Piotr},
  title = {Sequencing BPS spectra},
  journal = {JHEP},
  volume = {03},
  pages = {004},
  year = {2016},
  doi = {10.1007/JHEP03(2016)004},
  eprint = {1512.07883},
  archivePrefix = {arXiv},
  primaryClass = {hep-th}
}

@article{Labastida:1995ki,
  author = {Labastida, J. M. F. and Perez, E.},
  title = {A Relation Between the Kauffman and the HOMFLY Polynomials for Torus Knots},
  journal = {Journal of Mathematical Physics},
  volume = {37},
  year = {1996},
  pages = {2013--2043},
  eprint = {q-alg/9507031},
  archivePrefix = {arXiv}
}

\end{document}